\documentclass{aa}  

\usepackage{graphicx}
\usepackage{natbib}
\usepackage{txfonts}
\usepackage[fleqn]{amsmath}
\newcommand{\beq}{\begin{equation}}
\newcommand{\eeq}{\end{equation}}
\newcommand{\bea}{\begin{eqnarray}}
\newcommand{\eea}{\end{eqnarray}}
\newcommand{\rem}[1]{ }

\def\cgs{erg~cm$^{-2}$~s$^{-1}$}
\def\cm{cm$^{-2}$}
\def\ergs{erg~s$^{-1}$}
\def\kms{km~s$^{-1}$}
\def\whz{W~Hz$^{-1}$}

\def\sfr{$M_{\odot}$~yr$^{-1}$}
\def\msun{$M_{\odot}$}

\def\lesssim{\mathrel{\hbox{\rlap{\hbox{\lower3pt\hbox{$\sim$}}}\hbox{\raise2pt\hbox{$<$}}}}}
\def\gtrsim{\mathrel{\hbox{\rlap{\hbox{\lower3pt\hbox{$\sim$}}}\hbox{\raise2pt\hbox{$>$}}}}}

\begin{document} 

\title{Discovery of a galaxy overdensity around a powerful, heavily obscured FRII radio galaxy at z=1.7: star formation promoted by large-scale AGN feedback?}

   \author{
   R.~Gilli\inst{1},
   M.~Mignoli\inst{1},
   A.~Peca\inst{1},
   R.~Nanni\inst{1,2},
   I.~Prandoni\inst{3},
   E.~Liuzzo\inst{3},
   Q.~D'Amato\inst{2,3},      
   M.~Brusa\inst{2},
   F.~Calura\inst{1},
   G.B.~Caminha\inst{4},
   M.~Chiaberge\inst{5},
   A.~Comastri\inst{1},
   O.~Cucciati\inst{1},
   F.~Cusano\inst{1},
   P.~Grandi\inst{1},
   R.~Decarli\inst{1},  
   G.~Lanzuisi\inst{1},
   F.~Mannucci\inst{6},
   E. Pinna\inst{6},
   P.~Tozzi\inst{6},
   E.~Vanzella\inst{1},
   C.~Vignali\inst{2,1},
   F.~Vito\inst{7,8},
   B.~Balmaverde\inst{9},
   A.~Citro\inst{10},
   N.~Cappelluti\inst{11}
   G.~Zamorani\inst{1}
    \and
    C.~Norman\inst{5,12}
          }

   \institute{INAF -- Osservatorio di Astrofisica e Scienza dello Spazio di Bologna, Via P. Gobetti 93/3, 40129 Bologna, Italy\\
              \email{roberto.gilli@inaf.it}
         \and
         Dipartimento di Fisica e Astronomia, Universit\`a degli Studi di Bologna, Via P. Gobetti 93/2, 40129 Bologna, Italy
        \and
        INAF -- Istituto di Radioastronomia, Via P. Gobetti 101, 40129 Bologna, Italy
        \and
        Kapteyn Astronomical Institute, University of Groningen, Postbus 800, 9700 AV Groningen, The Netherlands
        \and
        Space Telescope Science Institute, 3700 San Martin Drive, Baltimore, MD 21218, USA
        \and
        INAF -- Osservatorio Astrofisico di Arcetri, Largo Enrico Fermi 5, 50125 Firenze, Italy
        \and
        Instituto de Astrofisica and Centro de  Astroingenieria, Facultad de Fisica, Pontificia Universidad Catolica de Chile, Casilla 306, Santiago 22, Chile
        \and
        Chinese Academy of Sciences South America Center for Astronomy, National Astronomical Observatories, CAS, Beijing 100012, China
        \and
        INAF - Osservatorio Astrofisico di Torino, Via Osservatorio 20, 10025 Pino Torinese, Italy
        \and
        Center for Gravitation, Cosmology and Astrophysics, Department of Physics, University of Wisconsin-Milwaukee, 3135 N. Maryland Avenue, Milwaukee, WI 53211, USA
        \and
        Physics Department, University of Miami, Coral Gables, FL, 33124, USA
        \and
        Department of Physics and Astronomy, Johns Hopkins University, Baltimore, MD 21218, USA}


\titlerunning{R. Gilli et al.}
\authorrunning{Discovery of a z=1.7 galaxy overdensity around a powerful, heavily obscured FRII}
  \abstract
{We report the discovery of a galaxy overdensity around a Compton-thick Fanaroff-Riley type II (FRII) radio galaxy at z=1.7 in the deep multiband survey around
the z=6.3 quasi-stellar object (QSO) SDSS J1030+0524. Based on a 6hr VLT/MUSE and on a 4hr LBT/LUCI observation, we identify at least eight galaxy members in this structure with spectroscopic redshift z=1.687-1.699, including the FRII galaxy at z=1.699. Most members are distributed within 400 kpc from the FRII core.  Nonetheless, the whole structure is likely much more extended, as one of the members was serendipitously found at $\sim$800 kpc projected separation. The classic radio structure of the FRII itself extends for $\sim$ 600 kpc across the sky. Most of the identified 
overdensity members are blue, compact galaxies that are actively forming stars at rates of $\sim$8-60 $M_{\odot}$~yr$^{-1}$. For the brightest of them, a half-light radius of 2.2$\pm$0.8 kpc at 8000\AA\ rest-frame was determined based on adaptive optics-assisted observations with LBT/SOUL in the Ks band. We do not observe any strong galaxy morphological segregation or concentration around the FRII core. This suggests that the structure is far from being virialized and likely constitutes the progenitor of a local massive galaxy group or cluster caught in its main assembly phase.
Based on a 500ks Chandra ACIS-I observation, we found that the FRII nucleus hosts a luminous QSO ($L_{2-10keV}=1.3\times10^{44}$ erg~s$^{-1}$, intrinsic and rest-frame) that is obscured by Compton-thick absorption ($N_H=1.5\pm0.6\times 10^{24}$\cm). Under standard bolometric corrections, the total measured radiative power ($L_{rad}\sim4 \times 10^{45}$ \ergs) is similar to the jet kinetic power that we estimated from radio observations at 150MHz ($P_{kin}=6.3 \times 10^{45}$ \ergs), in agreement with what is observed in powerful jetted AGN.

Our Chandra observation is the deepest so far for a distant FRII within a galaxy overdensity. It revealed significant diffuse X-ray emission within the region that is covered by the overdensity.
In particular, X-ray emission extending for $\sim$240 kpc is found around the eastern lobe of the FRII. Four out of the six MUSE star-forming galaxies in the overdensity
are distributed in an arc-like shape at the edge of this diffuse X-ray emission. These objects are concentrated within 200 kpc in the plane of the sky and within 450 kpc in radial separation. Three of them are even more concentrated and fall within 60 kpc in both transverse and radial distance. The probability of observing four out of the six $z=1.7$ sources by chance at the edge of the diffuse emission is negligible. In addition, these four galaxies have the highest specific star formation rates of the MUSE galaxies in the overdensity and lie above the main sequence of field galaxies of equal stellar mass at z=1.7. 

We propose that the diffuse X-rays originate from an expanding bubble of gas that is shock heated by the FRII jet, and that star formation is promoted by the compression of the cold interstellar medium of the galaxies around the bubble, which may be remarkable evidence of positive AGN feedback on cosmological scales.
We emphasize that our conclusions about the feedback are robust because even assuming that the diffuse X-ray emission arises from inverse Compton scattering of photons of the cosmic microwave backround by the relativistic electrons in the radio lobe, star formation may be promoted by the nonthermal pressure of the expanding lobe.
}


   \maketitle 
\section{Introduction}

Distant ($z\gtrsim1.5$) protoclusters and large-scale structures are ideal laboratories for investigating the complex processes that led to the assembly of local galaxy clusters. These processes involve mergers and interactions between gas-rich galaxies, fueling and growth of black holes at galaxy centers, and finally the (either positive or negative) AGN feedback on the intracluster medium (ICM) and on the star formation of member galaxies (see \citealt{overzier16} for a recent review). 

Gas-rich galaxies have been observed in high-z clusters and protoclusters \citep{hayashi17,noble17}, with star formation rates up to $800\;M_{\odot}$ yr$^{-1}$ \citep{santos15}. A large reservoir of diffuse cold and metal-rich molecular gas ($M_{H_2}\sim10^{11}\;M_{\odot}$) 
extending for 50-70 kpc was found around the radio galaxy at the center of the z=2.2 Spiderweb protocluster \citep{emonts18}. Its physical properties indicate that this gas is the product of mixing of large-scale gas outflows ejected by supernova (SN) winds and/or AGN activity, and that it is the seed of further star formation (see also \citealt{webb17} for another example of a large reservoir of molecular gas at the center of a z=1.7 cluster). 
All of the above suggests that the main transition from active to passively evolving galaxies in large-scale structures in fact occurs at  $z\sim1.5-2.0$, which is then a crucial epoch in the formation history of local massive clusters \citep{overzier16}.

High-z radio galaxies (HzRGs) are known to be excellent tracers of protoclusters and overdense environments \citep{pentericci00, miley08, chiaberge10}, and several HzRGs were found in which star formation in the host is triggered by the radio jet \citep{dey97,bicknell00}. Whether these powerful jets can also trigger star formation in companion galaxies remains an open question, and although numerical simulations indicate this as an efficient mechanism to form stars \citep{fragile17}, it has been observed in only a few systems \citep{croft06}. 

We report here the discovery of a large-scale structure at z$\sim$1.7 around a powerful Faranoff-Riley type II (FRII) radio galaxy in the field of the luminous z=6.3 quasar SDSS~J1030+0524. The presence of a powerful radio galaxy was revealed in 2003 by means of an observation of the quasar field at 1.4GHz with the Very Large Array (VLA) \citep{petric03}. \citet{nanni18} reanalyzed these data and measured the
flux, morphology, and extension of the radio galaxy. The object displays a classic FR II morphology, with an unresolved core (at 1.5" angular resolution), a jet pointing eastward, and two bright lobes (the west lobe is $>$6 times brighter than the east lobe), extending for 1.2 arcmin in total. In this paper we present the spectroscopic redshift measurement of the FRII host and the discovery of nearby galaxies that form a large-scale structure around it, as well as Chandra X-ray observations of the whole structure.

The paper is structured as follows: In Section 2 we present the entire set of multiband data that are available to study the galaxy overdensity. 
In Section 3 we describe the reduction and analysis of the data obtained i) at the Large Binocular Telescope (LBT) with the LBT Utility Camera in the Infrared (LUCI) and the
Single conjugate adaptive Optics Upgrade for LBT (SOUL), ii) with the Multi Unit Spectroscopic Explorer (MUSE) at the Very Large Telescope (VLT), and iii)
with the 2x2 array of the Advanced CCD Imaging Spectrometer (ACIS-I) on board the Chandra X-ray Observatory. In Section 4 we present the results on the
structure of the overdensity, star formation of its member galaxies, and power and obscuration of the FRII nucleus. In Section 5 we discuss the total mass of the
overdensity, the presence of another radio source that is a candidate member, the origin and interpretation of the diffuse X-ray emission, and finally, the evidence of positive AGN feedback on the star formation of some overdensity members. Our conclusions are presented in Section 6. 

A concordance cosmology with $H_0=70$ km~s$^{-1}$~Mpc$^{-1}$, $\Omega_m=0.3$ , and $\Omega_\Lambda=0.7$, in agreement within the errors with the Planck 2015 results \citep{planck16}, and a \citet{salpeter55} initial mass function (IMF) are used throughout this paper. In the adopted cosmology, the angular scale at z=1.7 is 8.5 kpc/arcsec.

\section{Multiband survey in the J1030 field}

We have accumulated a rich dataset of deep-and-wide multiband observations in the field centered at \hbox{RA=10$^h$ 30$^m$ 27$^s$}
\hbox{DEC=+05$^\circ$24$^\prime$55$^{\prime\prime}$} (hereafter the J1030 field), 
as a result of the intensive follow-up of the $z=6.3$ quasar SDSS~J1030+0524, the first discovered at $z>6$ \citep{fan01}, and of its small-to-large scale environment, which features the best candidate overdensity of galaxies around a quasar at such redshifts \citep{stiavelli05,kim09,morselli14,balmaverde17}. Our team is leading a major observational effort in the J1030 field by collecting and reanalyzing multiband data from major international facilities. \citet{morselli14} presented optical imaging in the $r,i,z$ filters obtained at the LBT using the Large Binocular Camera (LBC). \citet{balmaverde17} presented near-IR imaging in the $Y$ and $J$ filters obtained at the Canada France Hawaii Telescope using WIRCam (CFHT/WIRCam). Representative limiting magnitudes ($5\sigma$, AB) of $27.5,25,24$ in $r,z,J$, respectively, were obtained in these observations.

The field is part of the Multiwavelength Yale-Chile survey  (MUSYC, \citealt{gawiser06}), which provides additional imaging in $UBVRIzJHK$ down to $B=26$ and $K=23$ AB, for instance (see also \citealt{quadri07}),
and has also been entirely observed by Spitzer IRAC down to 24.5 AB mag at 3.6 $\mu$m, for example (3$\sigma$; \citealt{annunziatella18}).
In 2017 we observed the field with Chandra ACIS-I for 500ks \citep{nanni18}, making this field the fourth deepest extragalactic X-ray survey to date, 
and at the same time, the deepest X-ray observation of both a $z\sim 6$ QSO and of a distant galaxy overdensity around a powerful FRII galaxy.

The central part of the field has also been observed with the Hubble Space Telescope (HST) Advanced Camera for Surveys (ACS) \citep{stiavelli05, kim09} and Wide Field Camera 3 (WCF3) (PI Simcoe, unpublished), and VLT/MUSE (see Section 3.2). The host of the FRII radio galaxy falls at only 40 arcsec southwest of the QSO (see Fig.~1), and together with the inner regions of the overdensity, has then been covered by most imaging data in the field.

Optical and near-IR spectroscopic follow-up of the sources in the field is being conducted through dedicated campaigns at the LBT (with both the multi-object optical spectrograph MODS and
the near-IR spectrograph LUCI),  Keck (DEIMOS), and VLT (FORS2) telescopes. Detailed information about the full multiband imaging and spectroscopic coverage of the J1030 field can be found at the survey website\footnote{\url{http://www.oabo.inaf.it/~LBTz6/1030}}.
\begin{figure*}[t]
\begin{center}
\includegraphics[angle=0, width=13cm]{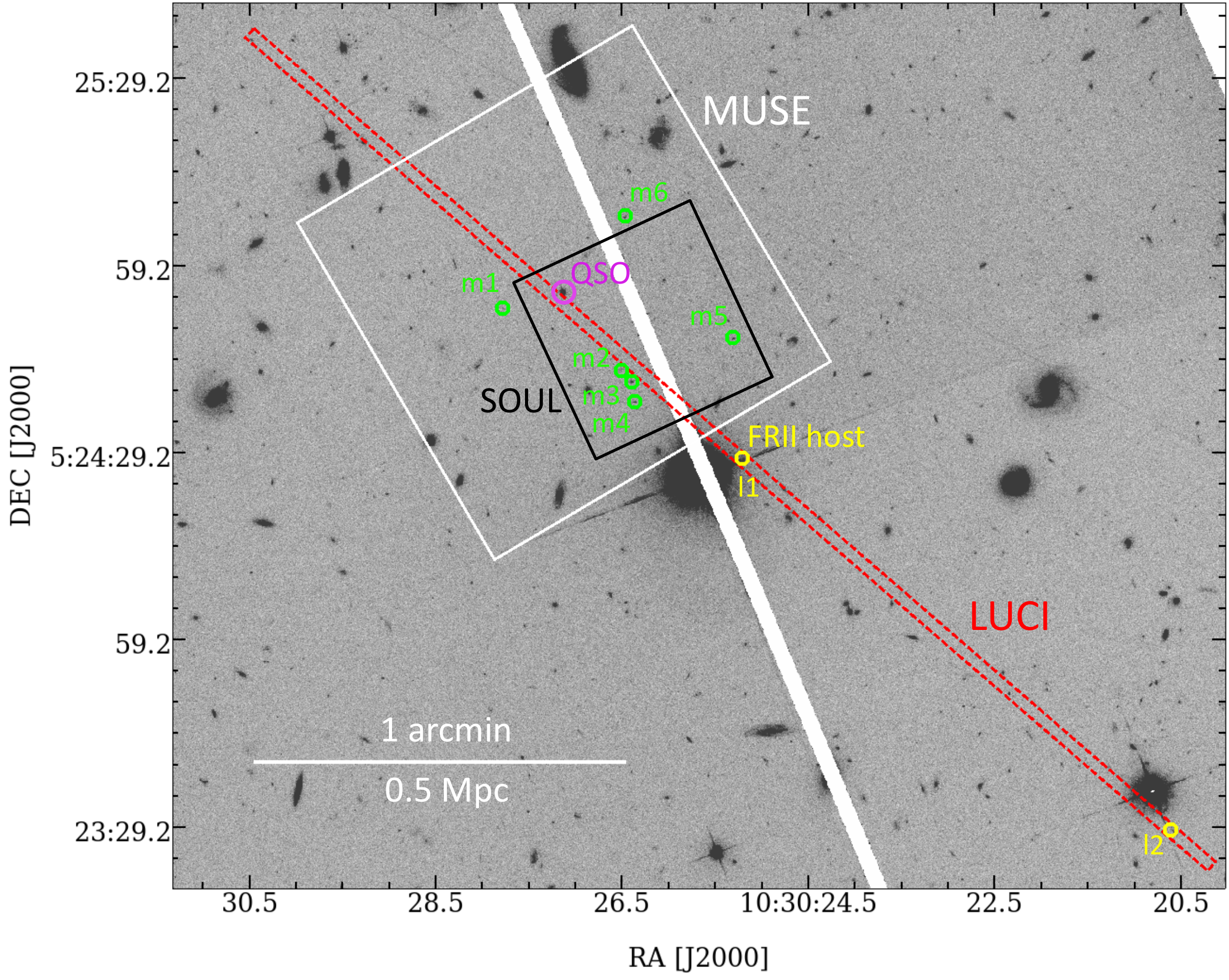}
\caption{HST/ACS F850LP image of the J1030 field (north is up and east is to the left). The white strip running across a very bright star is the gap between the two ACS CCDs.
The positions of the  LUCI long-slit (1"x205") observation of the FRII host and of the 60"x60" MUSE pointing are shown as a dashed red 
rectangle and a white square, respectively. The 30"x30" region observed with SOUL is shown as a black square. The yellow circles mark the FRII host at z=1.699 ($l1$) and a star-forming galaxy at z=1.697 ($l2$) that was found serendipitously along the LUCI slit. 
MUSE galaxies in the redshift range [1.687-1.697] are also shown as green circles and labeled ($m1-m6$). The position of the z=6.3 QSO SDSS~J1030+0524 is also marked in magenta.}
\label{full}
\end{center}
\end{figure*}

\section{Data reduction and analysis}

\subsection{LBT/LUCI}\label{luci}

The LBT/LUCI near-IR spectrum of the FRII radio galaxy at \hbox{RA=10$^h$ 30$^m$ 25.2$^s$}
\hbox{DEC=+05$^\circ$24$^\prime$28$^{\prime\prime}$} was obtained as part of the
INAF-LBT Strategic Program ID 2017/2018~\#18 (P.I. R.Gilli). This is a large
observing project devoted to spectroscopically follow up X-ray and radio sources in 
the J1030 field. The position of the FRII host in the J1030 field is shown in Fig.~\ref{full}. It falls close to a very bright star, but it is relatively
bright in the near-IR bands, and can therefore be accurately placed within a slit.

The FRII radio galaxy was observed during the nights of 2018 February 4 and 5 in the HKspec
($1.47-2.35\mu$m) band using the G200 grating. The total integration time was 4 hours, 
achieved through 72 individual exposures of 200s each, and the single exposures were
dithered along the slit. A slit of 1.0~arcsec$\times$ 3.4~arcmin was used, corresponding
to a spectral resolution of R$\approx$1000.
The LUCI data were reduced by the INAF LBT Spectroscopic Reduction Center in
Milan\footnote{http://www.iasf-milano.inaf.it/Research/lbt\_rg.html}.
The LBT spectroscopic pipeline was developed following the VIMOS experience
(Garilli et~al. 2012), but the reduction of LUCI spectroscopic data includes
new dedicated sky subtraction algorithms.

The analysis of the reduced 2D and 1D spectra of the radio galaxy is highlighted
by the clear presence of a bright emission line at $\lambda = 1.771\mu$m (see Fig.~\ref{luci}, $top$ panel). 
The spectral feature is a clear blend of two resolved emission lines, with an
intrinsic FWHM of $\approx$600~km~s$^{-1}$. The observed FWHM was deconvolved by subtracting in quadrature
the instrumental resolution as determined from adjacent sky lines. We identified the emission lines with the H$\alpha$
and [NII]6583\AA\ complex, and to measure the redshift, we performed a multiple
fit with three Gaussians. The relative intensity of the two lines of the
[NII]~doublet was fixed to the value of 3 \citep{acker89}, and the width of the [NII]
lines was matched to that of H$\alpha$. The continuum was fit with a linear
function in two spectral windows adjacent to the blended emission. No broad Balmer 
component is required by the fit, but the relatively low signal-to-noise ratio (S/N) of the spectrum
precludes any definitive conclusion regarding the presence of an additional
shallow broad component. The redshift obtained by the fit is 1.6987$\pm$0.0002. The measured flux and luminosity of the $H\alpha$ line are $2.2\times10^{-16}$\cgs and $4.4\times10^{42}$\ergs (with a $\sim14\%$ error), respectively. The measured value of the [NII]6583/H$\alpha$ ratio is $\sim$ 0.6-0.7, indicating, together
with the line FWHMs, that the FRII radio galaxy hosts an obscured (type~2) AGN \citep{cid11}.

Within the LUCI slit, at a distance of 1.5~arcmin southwest from the radio source (see Fig.~\ref{full}), a
serendipitous galaxy shows a spectrum with an unresolved emission line at 1.7695$\mu$m.
We identified the line as H$\alpha$ at $z=1.6966$. This redshift measurement is also supported by the detection
of a faint [NII]6583 emission feature. The measured flux and luminosity of the $H\alpha$ line are $2.4\times10^{-17}$\cgs and $4.8\times10^{41}$\ergs (with a $\sim25\%$ error), respectively.
The narrowness of the H$\alpha$ emission line, the low [NII]/H$\alpha$ flux ratio ($< 0.2$), and
the blue color of the galaxy are all suggestive of a young star-forming galaxy. Its near-IR spectrum
is shown in Fig.~\ref{luci} ({\it bottom panel}).

\begin{figure}[t]
\includegraphics[angle=0, width=9cm]{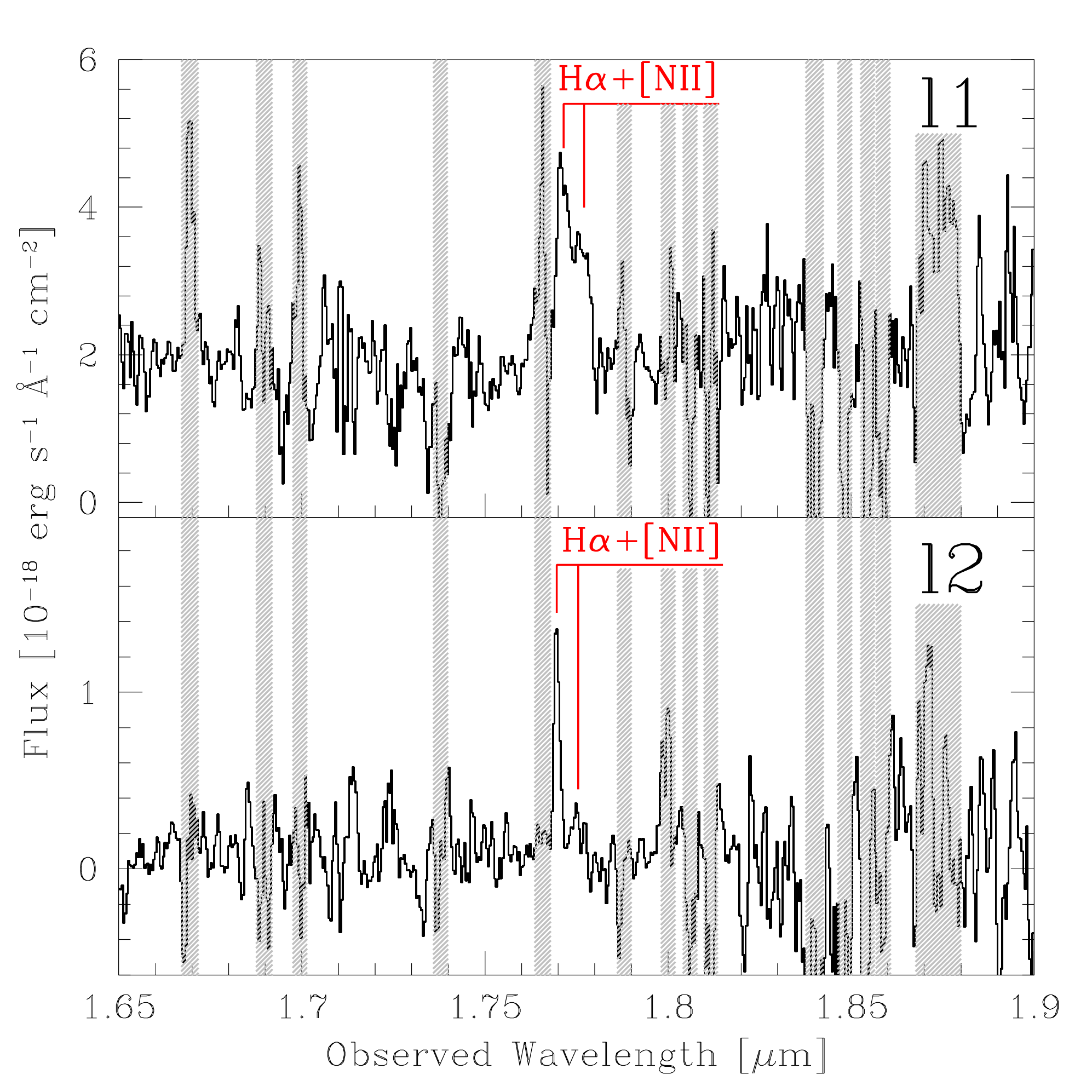}
\caption{$Top$: LBT/LUCI near-IR spectrum of the FRII host at z=1.6987. $Bottom$: LBT/LUCI near-IR spectrum of the serendipitous galaxy found at z=1.6966. H$\alpha$ and [NII] are marked in both panels. Gray bands show spectral regions with strong sky lines. } 
\label{luci}
\end{figure}

\begin{figure}[t]
\includegraphics[angle=0, width=9.cm]{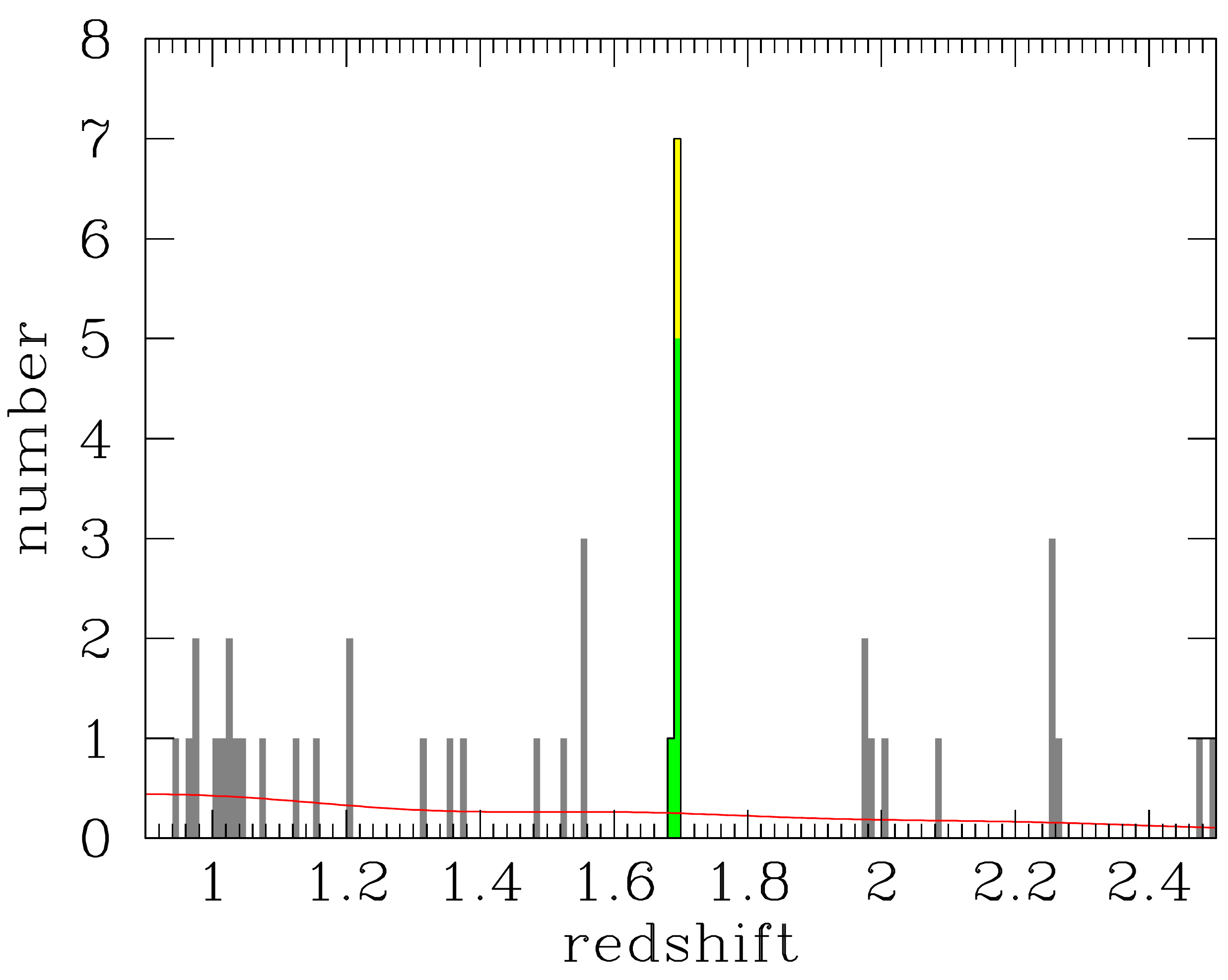}
\caption{Redshift distribution of MUSE sources at z=0.9-2.5 in bins of $\Delta z=0.01$ (gray histogram). The six MUSE sources ($m1-m6$) in the z=1.69 overdensity are shown
in light green. The red curve shows the expected background curve, obtained by smoothing the MUSE redshift distribution, used to quantify the significance of the redshift structure. 
The two additional sources at z=1.69 found by LUCI (including the FRII host) are shown in yellow.}
\label{zdist}
\end{figure}

\begin{figure}[t]
\includegraphics[angle=0, width=9cm]{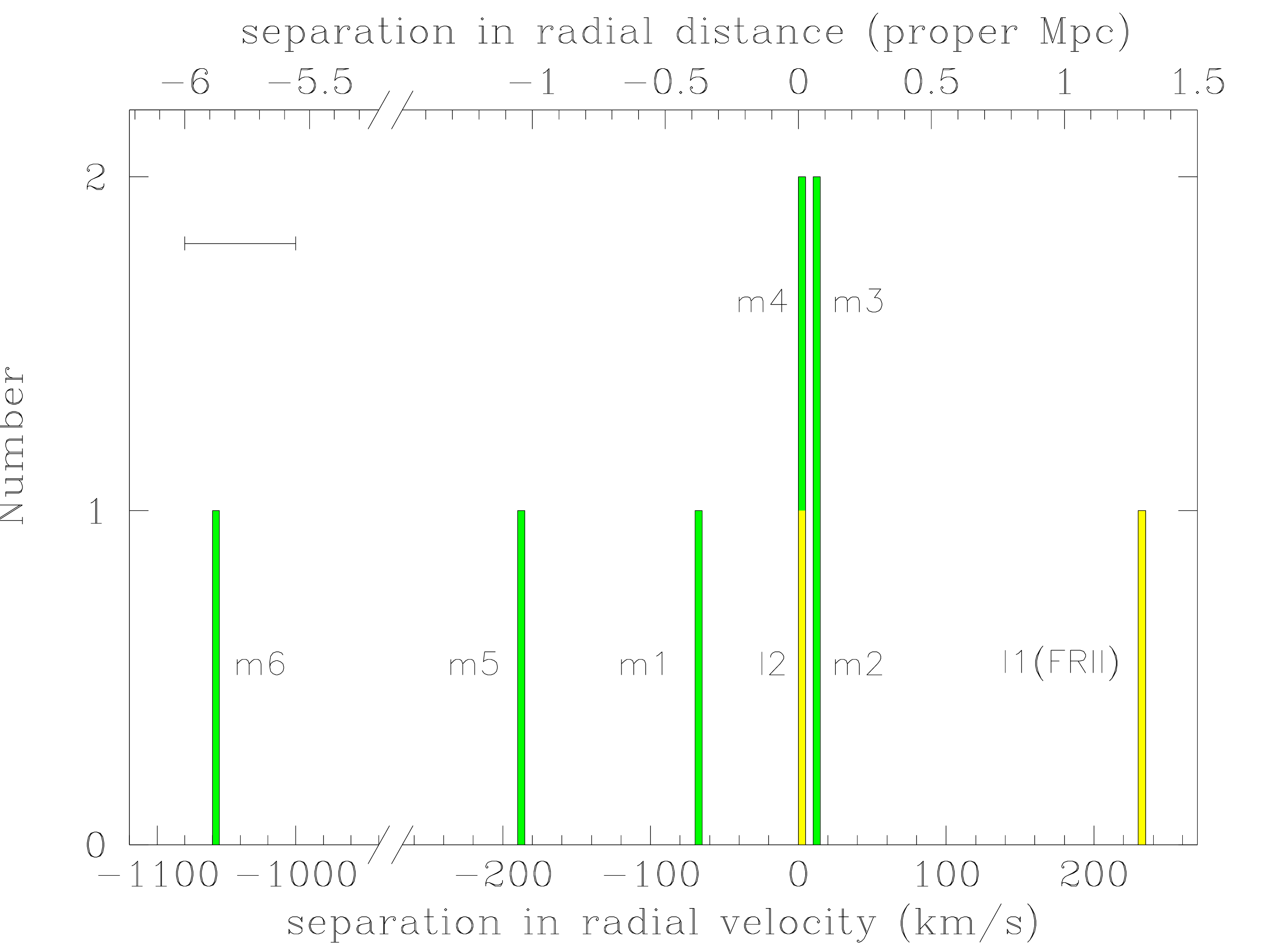}
\caption{Distribution of the eight overdensity members in rest-frame radial velocity space (lower x-axis) and in radial separations (upper x-axis), assuming the median redshift of the sample as the zero point. 
Radial separations are computed assuming that the overdensity members have negligible peculiar velocities. Velocity bins are 5 km/s wide. Green and yellow bars refer to redshifts measured by MUSE and LUCI, respectively, and the position of the FRII host is also labeled. The four MUSE sources in the arc around the diffuse X-ray emission ($m1-m4$) are all concentrated within 450 kpc radial, and the three of them that lie within 60 kpc on the plane of the sky ($m2-m4$), also lie within 60 kpc in the radial direction. The typical uncertainty introduced by redshift errors ($\Delta z \pm 0.0004$, see Table 1) is shown by the horizontal bar.}
\label{dv}
\end{figure}

\subsection{VLT/MUSE}

The field of SDSS J1030+0524 was observed by MUSE \citep{2010SPIE.7735E..08B, 2014Msngr.157...13B} between June and July 2016 
under the program ID 095.A-0714 (PI Karman) for a total of  $\approx 6.4$ hours of exposure time. 
The observations consist of a total of 16 target exposures with a small dither pattern and $90^{o}$ rotations to reduce instrumental features in the final stacked data.
We used the MUSE reduction pipeline version 1.6.2 \citep{2014ASPC..485..451W} to process the individual exposures and to create the final stacked data cube.
All usual calibration recipes (bias, flat field, wavelength, flux calibration, etc.) were applied to the raw exposures in order to create the corresponding data cubes and \text{PIXELTABLE}s.
We checked each single data cube but failed to find large differences between them.
The final data cube (with the full 6.4 hour depth) was created by combining the reduced \text{PIXELTABLE}s taking into account the small offsets between exposures.
In a final step we defined the sky regions from the white image and applied the Zurich atmosphere purge \citep[ZAP version 1.0,][]{2016MNRAS.458.3210S} to reduce sky residuals that were still present on the data.
The final data cube has a wavelength range of 4750$\AA$ to 9350$\AA$, a spectral sampling of 1.25\AA \  and a spatial sampling of $0\arcsec.2$, 
covering an area of 1 sq-arcmin (see Fig.~\ref{full}) centered at  \hbox{RA=10$^h$30$^m$27$^s$}
\hbox{DEC=+05$^\circ$24$^\prime$55$^{\prime \prime}$}, that is, the sky position of the QSO~SDSSJ1030+0524. 
The seeing conditions were very good, with an average seeing of 0.6 arcsec, as directly measured on bright stars in the MUSE data cube.

\begin{figure}[t]
\includegraphics[angle=0, width=9cm]{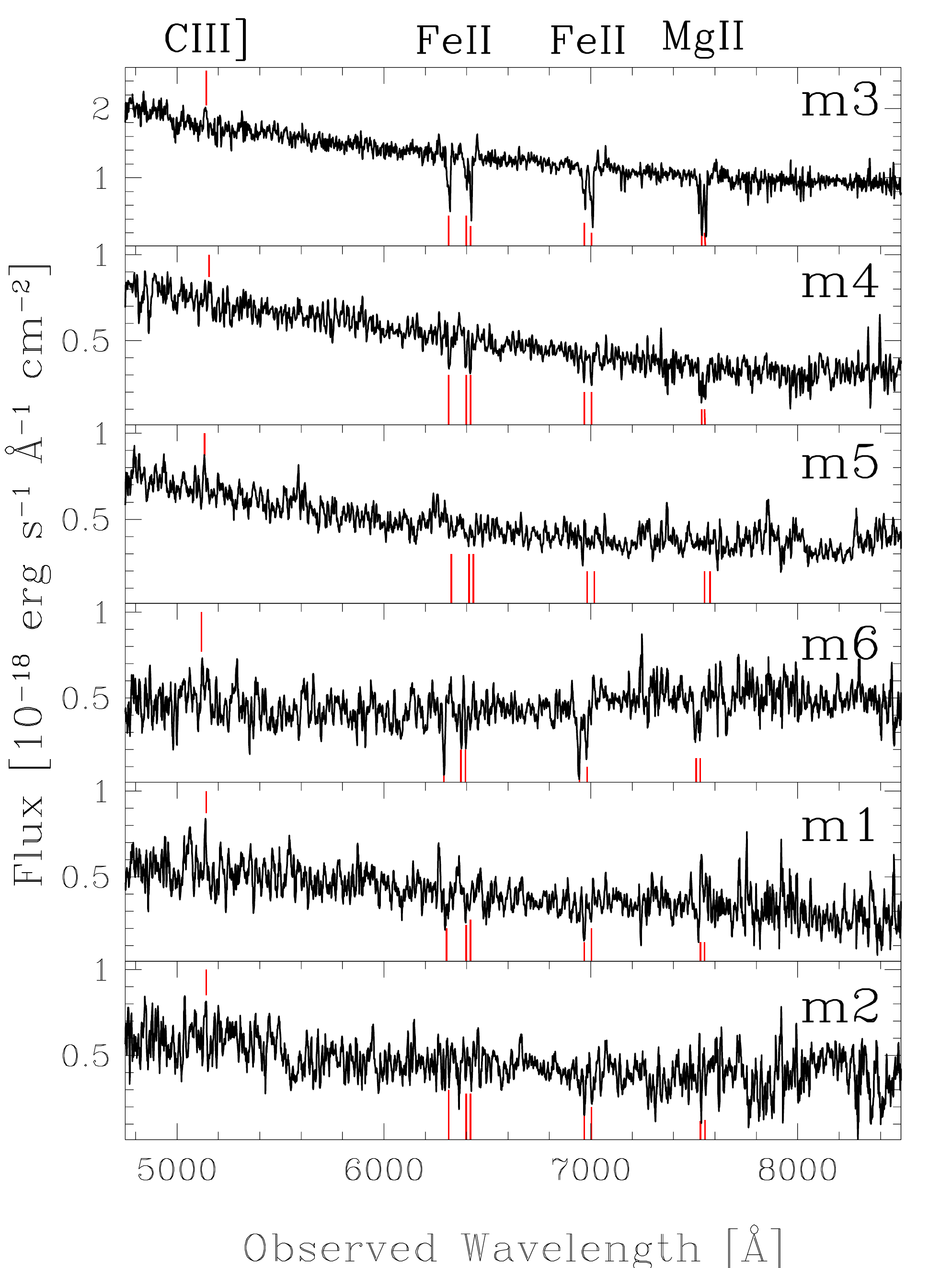}
\caption{MUSE spectra of the six star-forming galaxies in the $z=1.7$ overdensity sorted by decreasing UV flux. The main absorption and emission lines used in the redshift determination
are labeled.}
\label{muse}
\end{figure}

To identify the sources in the MUSE field of view (FoV), we used SExtractor (Bertin \& Arnouts 1996) on the white image, obtained by summing the flux at all wavelengths for each spaxel.
To avoid the loss of objects with extreme colors, we also ran SExtractor on sliced images in three different wavelength ranges (each 1500\AA\  wide). After combining the four lists and
performing a visual inspection to remove artifacts, we had a sample of 138 sources in the MUSE data cube, including the central quasi-stellar object (QSO). 
For each source we extracted a 1D spectrum by combining the spaxels inside a 3-pixel aperture that matches the seeing FWHM. We successfully measured
the redshift up to $z \gtrsim6$ for 87 of the 138 sources. As expected for such a small sky region, the redshift distribution is spiky, but a particularly striking feature is present at $z\sim1.7$, 
which is where the FRII host and the serendipitous source in the LUCI slit also lie (see Fig.~\ref{zdist}), with 5(6) MUSE galaxies within  $\Delta z<0.0018(0.096)$, that is, within 200(1060)~\kms (see also Fig.~\ref{dv}).
This is noteworthy because measuring redshifts with MUSE is
particularly difficult at this redshift owing to the absence of strong emission lines in the spectral range. For the two brightest galaxies ($m3$ and $m4$), the redshifts were measured
by finding seven interstellar absorption lines (produced by FeII and MgII transitions), computing redshifts for each identified line, and combining them into a single average value. 
For the other galaxies the redshift was obtained by cross-correlating their spectra with a star-forming template. 
Interstellar UV absorption lines are known to have velocity offsets with respect to the nebular emission lines
that trace the galaxy systemic velocities. To correct MUSE redshifts to the systemic redshifts and to make them comparable with the redshifts of LUCI spectra, we applied
a correction for interstellar gas outflows of $\Delta$v=135$\pm$22 \kms
(from \citealt{talia12}). In all of them the redshift is confirmed by the detection of the faint 
CIII]$\lambda$~1909 emission line. The MUSE spectra of the six galaxies in the $z=1.7$ spike are shown in Fig.~\ref{muse}.

\begin{table*}
\begin{center}
\begin{tabular}{cccccc}
\hline \hline
ID& RA& DEC& $K_{AB}$& $z_{spec}$& Notes\\
(1)& (2)& (3)& (4)& (5)& (6)\\  
\hline
$m1$& 10:30:27.73& +05:24:52.3& $23.97\pm0.45$& $1.6960\pm0.0005$& arc\\
$m2$& 10:30:26.46& +05:24:42.4& $23.14\pm0.23$& $1.6967\pm0.0004$& arc\\
$m3$& 10:30:26.34& +05:24:40.5& $22.21\pm0.14$& $1.6967\pm0.0002$& arc\\
$m4$& 10:30:26.31& +05:24:37.4& $22.67\pm0.28$& $1.6966\pm0.0003$& arc\\
$m5$& 10:30:25.26& +05:24:47.6& $23.34\pm0.34$& $1.6949\pm0.0004$& ---\\
$m6$& 10:30:26.42& +05:25:07.1& $22.12\pm0.21$& $1.6871\pm0.0003$& ---\\
$l1$&   10:30:25.20& +05:24:28.4& $20.90\pm0.12$& $1.6987\pm0.0002$& FRII\\
$l2$&   10:30:20.56& +05:23:28.7& $22.92\pm0.30$& $1.6966\pm0.0004$& ---\\
\hline                                             
\end{tabular}
\caption{List of spectroscopically confirmed overdensity members. Columns: 1) Source+redshift identifier: $m*$ and $l*$ refer to measurements by MUSE and LUCI, respectively; 2) and 3) source coordinates from either MUSE or LBT/LBC z-band data; 4) K-band magnitude from the MUSYC-deep catalog \citep{quadri07}. For $m1$, $l1$ and $l2$ we manually performed K-band photometry on the MUSYC-deep image. 5) redshift and 1$\sigma$ error; 6) Notes: Galaxies distributed in the arc-like shape at the edge of component A of the diffuse X-ray emission (see text) are labeled "arc".} 
\label{sample}
\end{center}
\end{table*}

\subsection{LBT adaptive-optics observations with SOUL}\label{soul}

The J1030 fields was observed using the new adaptive-optics (AO) system SOUL on the LBT telescope \citep{pinna16}. This is an upgrade of the existing AO system (FLAO, \citealt{esposito11}) implementing a pyramid wavefront sensor for a natural guide star (NGS) and an adaptive secondary mirror. With respect to FLAO, SOUL allows for a better AO correction and the use of fainter NGS. Observations were obtained on 8 April 2019 during the commissioning of the SOUL system with the LUCI1 near-IR camera. The bright star (R$\sim$12) close to the FRII radio galaxy (see Fig.~\ref{full}) allows for high-resolution AO observations.  The field was observed for 40 min in the Ks filter under seeing between 0.8'' and 1.0'' FWHM, and the data were reduced with standard procedures.
Although the reference star was found to be a double system with 0.4'' of separation and a factor of $\sim$4 in flux ratio, the AO correction provided a point spread function (PSF) down to FWHM=72$\times$76 mas at 23'' of distance to the NGS on the single one-minute images. The final combined image has FWHM=90$\times$120 mas at the same distance. 

\begin{figure}
\includegraphics[angle=0, width=9cm]{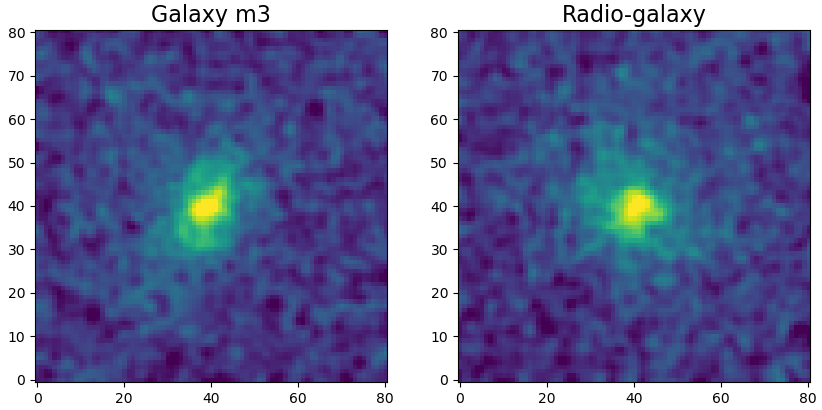}
\caption{Ks-band images of the galaxy $m3$ ($left$) and of the radio galaxy that is the candidate overdensity member discussed in Sect.~\ref{arinp} ($right$) obtained with the AO system SOUL at the LBT. Each cutout is 1.2''$\times$1.2'' , and the axes are in pixel units (the scale is 15 mas/pixel).}
\label{soul_image}
\end{figure}

One of the MUSE galaxies in the structure at $z=1.7$ ($m3$, see Fig.~\ref{full} and Table~\ref{sample}) and a radio galaxy that is also a candidate source at the
same redshift (see Section~\ref{arinp}) were detected with high S/N (see Fig.~\ref{soul_image}). The two galaxies are well resolved, but the depth of the image is not enough to perform a full PSF deconvolution and morphological fitting. Nevertheless, a reliable half-light radius $r_{50}$ can be measured by aperture photometry (circular apertures were used). For $m3$ we obtained $r_{50}=0.26\pm0.1$'', corresponding to 2.2$\pm$0.8 kpc at z$\sim$1.7, and for the radio galaxy  we obtained $r_{50}=0.27\pm0.05$'', corresponding to 2.3$\pm$0.4 kpc at z$\sim$1.7.

\subsection{Chandra/ACIS-I}

We observed the J1030 field with Chandra/ACIS-I for a total of 479ks. The observation was divided into ten different pointings with roughly the same aim point taken between January and May 2017. The data were taken in the {\sc vfaint} mode, processed using CIAO v4.8, and filtered using standard ASCA grades. The astrometry of each pointing was registered to a reference source catalog that we derived from CFHT/WIRCam observations in the Y and J bands \citep{balmaverde17}. More details on the Chandra data reduction are given in \citet{nanni18}. 
An X-ray source catalog is being derived from these observations that is based on CIAO {\sc wavdetect} \citep{freeman02} for source detection and on ACIS Extract \citep{broos10} for source photometry. 

\subsubsection{FRII nucleus}

\begin{figure}[t]
\includegraphics[angle=0, width=9cm]{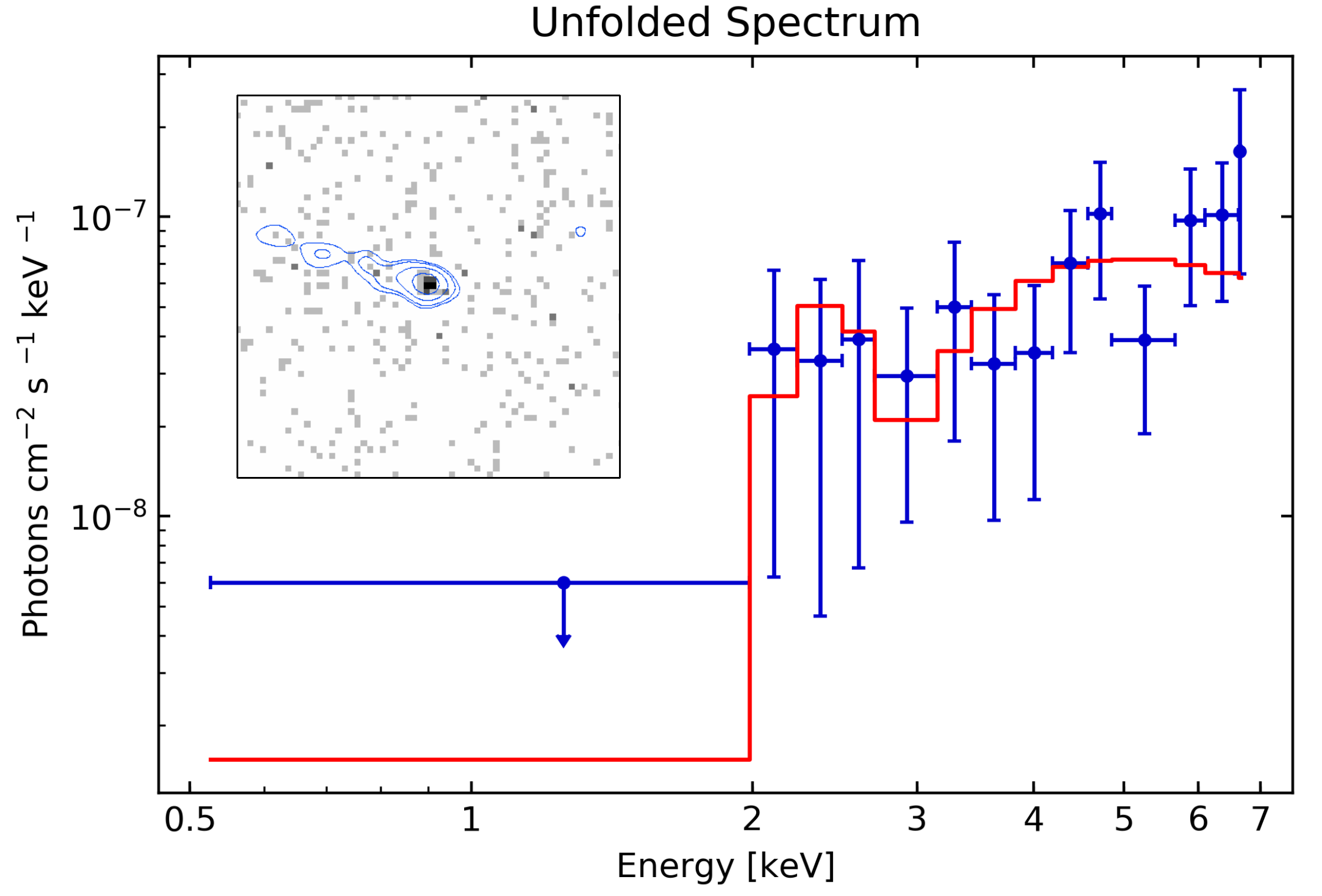}
\caption{Response-corrected Chandra/ACIS-I X-ray spectrum of the FRII nucleus (XID189) and best-fit model (in red). By fixing the photon index to 1.8, a best-fit column density of $N_H=1.5^{+0.6}_{-0.5}\times 10^{24}$ cm$^{-2}$ and an intrinsic deabsorbed luminosity in the 2-10 keV rest-frame of $L_X=1.3\times10^{44}$ erg~s$^{-1}$ are obtained, which qualify XID189 as a Compton-thick QSO. The spectrum
has been rebinned for display purposes. The inset shows a 30"x30" cutout of the 0.5-7 keV Chandra raw image around the source. VLA radio contours at 1.4GHz are overlaid in blue 
(with a $\sqrt 3 $ geometric progression starting at 30$\mu$Jy beam$^{-1}$).}
\label{xid189}
\end{figure}

The core of the FRII corresponds to the X-ray source XID189 in the catalog above. XID189 is only detected in the
hard 2-7 keV band ($>5\sigma$) with a flux of $f_{2-7}=2.2^{+0.3}_{-0.4}\times 10^{-15}$\cgs. We extracted the spectra of XID189 from each individual pointing and combined them using {\sc combine\_spectra} in CIAO. A circle of 1.5" radius was used as the source extraction region. A similar procedure was used to derive the background spectrum using an annulus around the source with inner
and outer radius of 3" and 6", respectively. The X-ray spectrum of XID189 was grouped to a minimum of one count per energy bin and then analyzed with XSPEC v12.5.3 using the C-statistic \citep{cash79} to estimate the best-fit parameters. All errors are given at the 1$\sigma$ level. A total of $31^{+7}_{-6}$ net counts were measured in the 0.5-7 keV range (1.3-19 keV rest-frame). We fit the spectrum using a simple absorbed power-law model. The absorption was modeled through XSPEC {\sc plcabs} \citep{yaqoob97}, which assumes an isotropic source of photons enshrouded in a spherical matter distribution. The advantage of this model with respect to the commonly used {\sc zphabs} or {\sc zwabs} is that in addition to photoelectric absorption, it correctly accounts for the effects of Compton scattering (at least up to column densities of $5\times 10^{24}$ cm$^{-2}$ and rest-frame energies of $\sim20$ keV). 
The use of more sophisticated absorption models such as {\sc mytorus} \citep{mytorus} would be inadequate for the low photon statistics measured in XID189. When we fixed the power-law photon index to 
$\Gamma=1.8$, we measured a column density of $N_H=1.5^{+0.6}_{-0.5}\times10^{24}$cm$^{-2}$ and absorption-corrected luminosity of $L_X=1.3\times10^{44}$\ergs in the 2-10 keV band rest-frame, which qualify XID189 as a heavily obscured, Compton-thick quasar. The Chandra image and spectrum of XID189 are shown in Fig.~\ref{xid189}.

\begin{figure*}[t]
\includegraphics[angle=0, width=11.1cm]{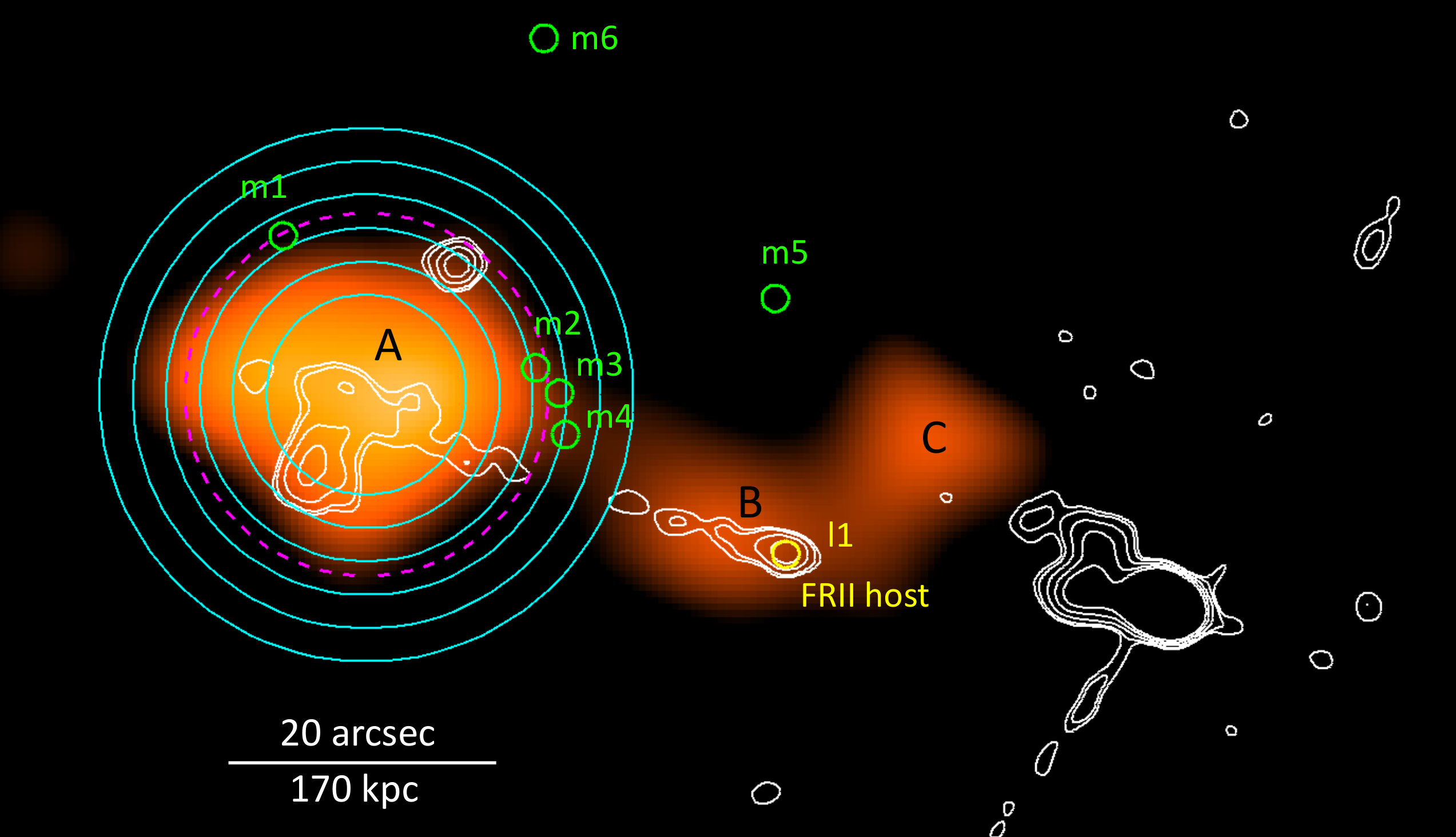}
\hspace{0.2cm}
\includegraphics[angle=0, width=6.8cm]{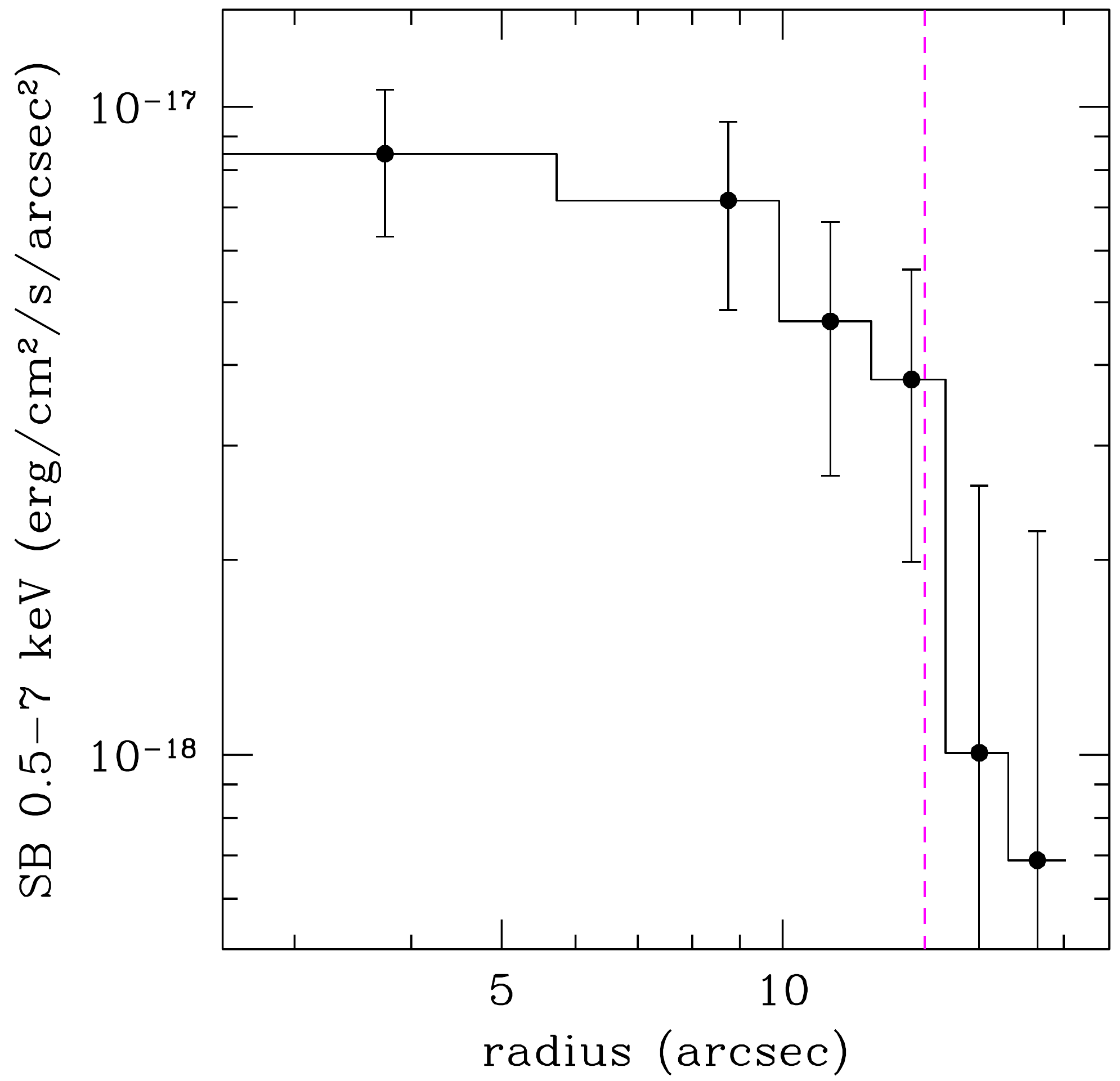}
\caption{$Left$: Point-source subtracted and smoothed Chandra/ACIS-I image in the 0.5-7 keV band. The main components A, B, and C of the diffuse X-ray emission are marked. Galaxies at z=1.69 are labeled as in Fig.~1 and Table~1 ($l2$ falls outside this image). 
VLA radio contours at 1.4 GHz are overplotted in white (same levels as in Fig.~\ref{xid189}): they show the full morphology (with jet and lobes) of the FRII radio galaxy and an additional radio source that is also possibly part of the overdensity (see text). The annuli used to
compute the surface brightness profile of component A (see $right$ panel) are shown in cyan. The dashed magenta circle (14" radius around the X-ray centroid) shows the location of the overdensity galaxies $m1$-$m4$. $Right$: Surface brightness profile of the diffuse component A as measured on the (point-source subtracted) Chandra 0.5-7 keV image {\it \textup{before smoothing}}, and using the extraction annuli shown in the $left$ panel. The radial profile is
background subtracted and a power-law spectrum with $\Gamma=1.6$ is assumed to convert count rates into fluxes. The radius
at which the $m1-m4$ galaxies are found is shown by the magenta vertical line. The X-ray surface brightness sharply  decreases at this radius, and is consistent with zero beyond it.}
\label{csmooth}
\end{figure*}

\subsubsection{Diffuse X-ray emission}

Our deep Chandra observation revealed several regions of significant diffuse X-ray emission within the area covered by the galaxy overdensity. 
In Fig.~\ref{csmooth} ($left$) we show a map of the extended X-ray emission obtained by first removing point-like X-ray sources from the Chandra 0.5-7 keV image, hence replenishing the "holes" in the image with photons extracted from local background regions (using the CIAO {\sc dmfilth} tool), and finally smoothing it with the {\sc csmooth} tool using smoothing scales up to 10 pixels ($\sim$5"). We remark that all measurements presented in the following were performed on the unsmoothed images. The background was evaluated in a rectangular source-free region of equal and uniform exposure north of the diffuse emission. Spectral analysis was performed with XSPEC v12.5.3 on spectra grouped to at least one count per bin and using the C-statistic. At least three spots of significant 
X-ray emission are seen, which are labeled A, B, and C in Fig.~\ref{csmooth} ($left$). The most prominent emission (component A) is detected with $S/N=5.5$ in the 0.5-2 keV band. It extends for  $\sim30"\times20"$ and overlaps with the eastern lobe of the FRII galaxy. Significant emission ($S/N\sim5$) is also found along the direction of the radio jet (component B in Fig.~\ref{csmooth} $left$). Finally, low-significance emission ($S/N\sim2.4$ and 3.4 in the full and soft band, respectively) is found at 20 arcsec northeast of the FRII western lobe (component C in Fig.~\ref{csmooth} $left$). Despite its low significance, we consider component C as real because it is also visible in the X-ray images that have independently been obtained with XMM-Newton (see, e.g., \citealt{nanni18}). 

We note that the significant detection of diffuse X-ray emission around such a distant FRII galaxy has been made possible by the exceptionally deep Chandra observation of the system. We inspected our Chandra data by cutting the exposure at 120ks and 200ks, performed aperture photometry of component A, and compared the results with those from the final exposure. Again, in the unsmoothed data, the S/N of component A in the 0.5-7 keV band increases from ~2.1, to 3.3 and 5.5 with increasing exposure. Even considering the brightest of
the diffuse X-ray components, this can therefore be detected significantly ($S/N>3$) only with Chandra exposures larger than $\sim$200ks.

\begin{figure}[t]
\includegraphics[angle=0, width=9cm]{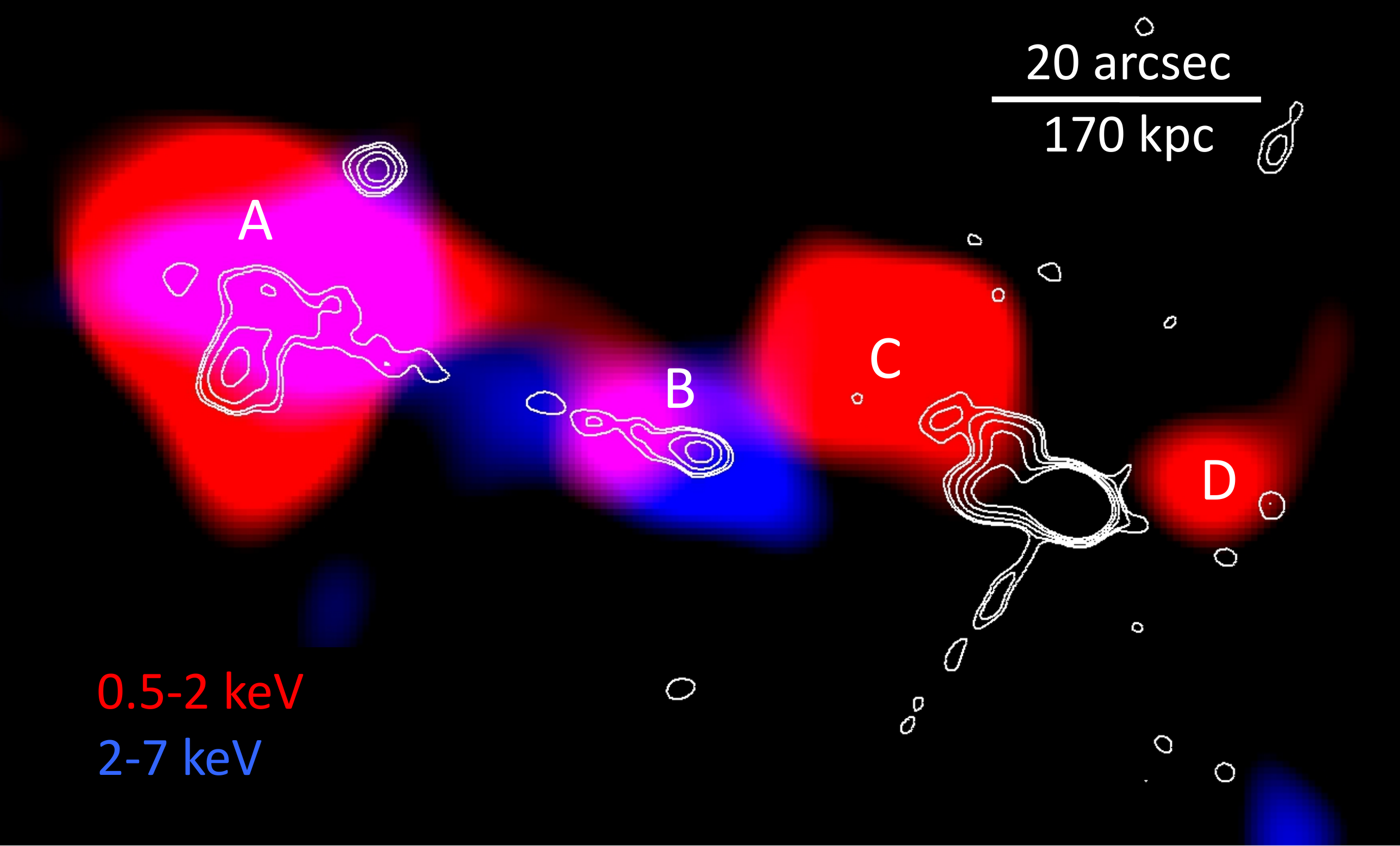}
\caption{Point-source subtracted and smoothed Chandra/ACIS-I X-ray color image of the diffuse emission (see text for details). Soft (0.5-2 keV) and hard (2-7 keV) X-rays are shown in red and blue, respectively. The X-ray emission is
shown down to a $\sim 2.5\sigma$ significance level. Radio contours are shown in white. }
\label{HR}
\end{figure}

We further investigated the diffuse X-ray emission in the soft and hard band separately, by smoothing point-like subtracted X-ray images following the procedure described above. In Fig.~\ref{HR} we show the overlap between the soft and hard diffuse X-ray emission. Another component (component D), emerges in the soft band only (with $S/N\sim2.5$). Inspection of the image reveals that hard X-ray photons are clearly associated with the radio emission pointing eastward of the FRII core and reaching component A, which is indeed globally harder than components C and D. The X-ray photometry of the four components is presented in Table~\ref{xdiffuse}.

\citet{nanni18} discussed the origin of component A and proposed that it can be either associated with feedback produced by the z=6.3 quasar (which is located in projection just above this structure) on its close environment, or with the eastern lobe and jet of the FRII 
radio galaxy. We extracted and analyzed the X-ray spectrum of component A ($\sim 100$ net counts in the full band), but could not identify any significant spectral feature to measure its redshift. 

Based on an in-depth analysis of the structure of the diffuse X-ray emission, we here suggest that this is likely to be produced at $z=1.7$. First, as discussed above and reported in Table~\ref{xdiffuse}, in addition to the main component A, we detected three more significant spots of diffuse X-ray emission across the whole structure of the FRII and of the related galaxy overdensity. 
The presence of multiple spots of diffuse X-ray radiation that are located in the same region covered by the FRII and by the overdensity structure provides a strong indication that diffuse X-rays are actually produced at $z=1.7$. Second, when X-ray point-like sources are removed from Chandra images, and, in particular, the nuclear emission of the z=6.3 QSO is removed, no significant diffuse X-ray emission reaches the position of the QSO (see, e.g., Fig.~\ref{sepia}). Despite some residual uncertainties that are related to the limited photon statistics, to the
point source subtraction process, and to the smoothing process, the lack of any significant diffuse X-ray structure directly emanating from the QSO location argues against an origin at $z=6.3$. In contrast, the eastern radio lobe of the FRII appears extremely well centered on component A, and it even bends southward after reaching the centroid of the diffuse X-ray emission: this also suggests some relation between the X-ray and radio data. Because of the above arguments we then assume here that most of the observed diffuse X-rays are produced at $z=1.7$.

We first fit component A with a thermal model ({\sc apec} in XSPEC) with metal abundances fixed to $0.3\times$ solar, as is typically measured in galaxy clusters \citep{balestra07}. The spectrum 
of component A is rather hard (see the photometry in Table~\ref{xdiffuse}), and only a lower bound to the gas temperature can be obtained  ($T\gtrsim5$ keV at 2$\sigma$). The total X-ray luminosity is $L_{2-10}\sim4\times10^{43}$\ergs\ in the 2-10 keV rest-frame.

Components C and D are instead softer, and in fact are not detected in the hard band. Their emission ($\sim21$ and $\sim 12$ net counts, respectively) is well fit by an {\sc apec} model with $T\approx 1$keV (see Table~\ref{xdiffuse}; a fit with a power law returns implausibly steep photon indices, $\Gamma\sim4-5$). The 2-10 keV rest-frame luminosity of components C and D is $L_{2-10}\sim1.3\times10^{43}$\ergs 
and $\sim6\times10^{42}$\ergs, respectively.

The hard spectrum of component A might indicate that nonthermal processes such as synchrotron radiation or inverse Compton 
scattering of cosmic microwave background (CMB) photons (IC-CMB) by the relativistic electrons in the lobe provide a non-negligible contribution to its total emission (e.g., \citealt{smail12}).
This hypothesis is discussed in Section~\ref{diffx}.

As shown in Fig~\ref{HR}, the soft X-ray emission seen in component A appears to be more extended than the hard emission, supporting the idea that we
may be seeing a mixture of IC-CMB emission in its center, and softer thermal emission on larger scales.  
In this case, the actual values of the gas temperature and luminosity of component A may be lower than those reported in Table~2.
The origin of the diffuse X-ray emission, especially in component A, is discussed in more detail in Sections \ref{diffx} and \ref{feedback}.

We finally inspected component B, which coincides with the radio jet emission and features hard X-ray emission, as is readily apparent
from Fig.~\ref{HR}. It contains $\sim 60$ net counts in the 0.5-7 keV band, and $\sim80\%$ of them
are at $E>2$ keV. A power-law fit returns a flat photon index $\Gamma=0.1\pm0.4$ and a flux of $f_{0.5-7}=2.9\times 10^{-15}$\cgs, corresponding to a 2-10 keV rest-frame luminosity of $L_{2-10}\sim1.6\times10^{43}$\ergs.

\begin{table*}
\begin{tabular}{cccccccccccc}
\hline \hline
ID& RA& DEC& \multicolumn{3}{c}{Net counts}& R& $\Gamma$& $kT$& $f_{0.5-7}$& $L_{2-10}$& $C/dof$\\
& & & full& soft& hard& ($^{\prime \prime}$)& &(keV)& ($10^{-15}$cgs)& $(10^{43}$\ergs)& \\
(1)&(2)&(3)&(4)&(5)&(6)&(7)&(8)&(9)&(10)&(11)&(12)\\
\hline
A& 10:30:27.4& +05:24:39.5& 110$\pm$20& 62$\pm$12& 48$\pm$17& 14& $1.64_{-0.35}^{+0.39}$& -& 3.7&  4.3& 210.7/228 \\
  &             "      &                "          &    "         &       "          &       "           &    "&                          -&  $> 5$& 3.6& 4.4& 211.5/228\\
B& 10:30:25.5& +05:24:29.4&   61$\pm$12& 11$\pm$  6& 50$\pm$11&   8& $0.06_{-0.44}^{+0.36}$& -& 2.9&  1.6&   81.5/120\\
C& 10:30:24.4& +05:24:37.1&   24$\pm$10& 21$\pm$  6&    3$\pm7$& 10& -& $0.63_{-0.17}^{+0.28}$& 2.1&  1.3&  14.3/29\\
D& 10:30:22.7& +05:24:27.9&   12$\pm$8  & 12$\pm$  5&    0$\pm6$&   6& -& $0.78_{-0.20}^{+0.84}$& 0.8&  0.6&  20.4/20\\
\hline                                             
\end{tabular}
\caption{List of the diffuse X-ray emission components. Columns: 1) component identifier; 2) and 3) 0.5-7 keV centroid coordinates; 4), 5), 6) net counts in the 0.5-7 keV, 0.5-2 keV and 2-7 keV band, respectively; 7) extraction radius adopted for the photometric and spectral analysis; 8) best-fit photon index obtained with a power-law model (see text for details); 9) best-fit temperature obtained with a thermal model (see text for details); 10) 0.5-7 keV observed flux; 11) 2-10 keV rest-frame luminosity; and 12) best-fit statistics over degrees of freedom. All errors are given at the $1\sigma$ level.} 
\label{xdiffuse}
\label{xtab}
\end{table*}


\section{Results}

\subsection{Structure of the overdensity}

The analysis of LBT/LUCI and VLT/MUSE data allowed us to confirm eight objects in the redshift interval $z=1.687-1.699$, including the FRII host at z=1.699 (see Table~\ref{sample}).
Six of these objects are star-forming galaxies discovered by MUSE. The redshift distribution of the MUSE sources in the $z=0.9-2.5$ interval is shown in Fig.~\ref{zdist}. Following a similar approach to that described in \citet{gilli03}, we quantified the significance of the $z\sim1.7$ redshift structure by smoothing the redshift distribution of all galaxies in the MUSE field (with a Gaussian with $\sigma_z=0.2$) and considering this as the MUSE background redshift distribution (red curve in Fig.~\ref{zdist}). The Poisson probability of observing $N_g$=6 galaxies when $N_{bkg}$=0.26 are expected within $\Delta z<0.01$ is lower than $3.5\times10^{-7}$. 
As cosmic variance might be an issue on small fields such as the one studied here \citep{moster11}, we verified whether smoothing the observed redshift distribution provides a biased estimate of the shape and normalization of the “true” background galaxy distribution, which in turn would alter the significance of our measurement. 
We then downloaded the spectroscopic catalog obtained from the 3’x3’ MUSE observation of the Hubble Ultra Deep Field (HUDF,  \citealt{beckwith06}) consisting of nine MUSE pointings of 10 hr each \citep{bacon17,inami17}, and smoothed the MUSE HUDF redshift distribution to obtain a more accurate estimate of the average background shape. This curve was then renormalized to the average number of galaxies expected in a single MUSE pointing of the HUDF. We note that if anything, this procedure may somewhat overestimate the average galaxy background density because the HUDF observations are deeper than those in the J1030 field. The derived significance of the z=1.7 overdensity may then be regarded as conservative. The obtained distribution was found to be very similar to that shown in Fig. 3, and our estimates of the level and significance of our z=1.7 overdensity are therefore strengthened by adopting a more accurate background distribution. 

The structure at $z\sim1.7$ is highly significant, and corresponds to an overdensity of 
$\delta_g=N_g/N_{bkg}-1 = 22$. This in turn indicates that the FRII radio galaxy is the signpost of a high-z large-scale structure (see Section~\ref{mass} for an estimate of the structure mass). Most overdensity members lie at $<0.4$ Mpc projected separations from the FRII host, i.e. where MUSE 
data are available. However, the serendipitous detection of another overdensity member in the LUCI slit at 0.8 Mpc separation from the FRII host in the opposite direction suggests that the whole structure 
extends for at least 1.2 projected Mpc. The existence of other candidate members at 0.4-0.7 Mpc projected separation 
from the FRII host is indicated by the joint X-ray spectral and photometric redshift analysis of Chandra sources in the field (Peca et al. in preparation). The actual size of the structure can then only be established by further spectroscopic observations. 

In Fig.~\ref{dv} we show the distribution of the eight overdensity members in (rest-frame) velocity space. The velocity offset is computed with respect to the median redshift of the sources. 
All objects lie within $\Delta$v=1300~km~s$^{-1}$, and their velocity dispersion (computed with the gapper method, \citealt{beers90})  is  $325\pm 226$ \kms, which decreases to only 
$121\pm68$ \kms \ if the object at lower redshift  ($m6$ at z=1.6871, $\Delta$v$\sim -1060$ \kms in Fig.~\ref{dv}) is neglected. When only the six MUSE sources are considered, the velocity dispersion is 
$355\pm300$ \kms, which decreases to $85\pm56$ \kms\ when $m6$ is removed. These values are significantly lower than what is measured for massive galaxy clusters (e.g., $\sim 1000$ \kms\ 
for $M\sim10^{15}\;M_{\odot}$, \citealt{rbn02}) and similar to what is measured in low-mass groups with $M\sim10^{13}\;M_{\odot}$ \citep{mulchaey00}. In Section~\ref{mass} we derive a lower limit to the total mass of our system of $M\sim 1.5-2\times 10^{13}\;M_{\odot}$. As discussed above, the size, and 
hence the mass, of the structure is likely much larger than what can be estimated based on the MUSE data alone. The measured low-velocity dispersion  may then indicate that the system is still far from virialization. By assuming that peculiar velocities are negligible, 
we therefore converted the redshift differences among the overdensity members into radial separations, which are also reported in Fig.~\ref{dv}. All objects fall within a radial separation of $\sim 7$ physical Mpc.
Also, the four objects $m1-m4$ at the boundary of component A of the diffuse X-ray emission lie within $\Delta z$=0.0007, that is, within 450 radial kpc, and the three of them falling within 60 kpc in the plane of the sky ($m2-m4$, see Fig.~\ref{csmooth} $left$) 
also lie within $\Delta z$=0.0001, that is, within 63 kpc in the radial direction. There are obvious uncertainties in this estimate. On the one hand, if the structure is caught far from virialization and 
at the beginning of its collapse phase, coherent galaxy motions may lead to an underestimate of its true radial dimension. On the other hand, errors in the redshift measurement likely inflate the true radial separations. With these caveats in mind, we note that the radial separation of the galaxies falling at the edge of component A of the diffuse X-ray emission is comparable with the transverse dimension of the latter.

\subsection{Star formation, ages, and masses of the overdensity members}\label{sfragesmasses}

We computed the extinction-corrected star formation rate of the MUSE overdensity members $m1-m6$ starting from their absolute UV magnitude at 2800\AA\ rest-frame $M_{2800}$. To determine $M_{2800}$, we integrated the flux of their VLT/MUSE spectra in a bandwidth of 200~\AA\ centered on 2800~\AA. The absolute flux calibration of the MUSE data is very good \citep{kreckel17}, and the absence of slit losses allows us to confidently measure the  total UV flux emerging from the galaxies (as also demonstrated by the excellent agreement between the optical magnitudes obtained from broad-band photometry and the fluxes measured by MUSE at different wavelengths).
 To convert the far-UV (FUV) luminosity into the ongoing star formation rate (SFR) we used the conversion factor $K_{\rm FUV}\equiv {\rm SFR}/L_\nu{\rm(FUV)}=1.3\times10^{-28}$ for Z=Z$_\odot$ \citep{md14}, 
 where the SFR is expressed in units of \sfr and $L_\nu{\rm(FUV)}$ in units of \ergs~Hz$^{-1}$. The FUV continuum slope was measured in the range 1920$-$2175~\AA\  rest-frame, that is,
 5175$-$5865~\AA\ observed frame, which is relatively free from intense sky lines.
To transform the continuum slope parameter measured in this range into the classical UV continuum slope $\beta$ (where $f_{\lambda}\propto \lambda^\beta$), we adopted the relations
proposed by \citet{noll05}. Finally, the intrinsic SFR was computed taking into account the dust-extinction correction adopting the definition 
log(SFR$_{tot}$) = log(SFR$_{UV}$)+ 0.4 $\times A_{IRX}$ \citep{nordon13}, where SFR$_{\rm UV}$ is the uncorrected star formation rate and A$_{\rm IRX}$ is the effective UV attenuation derived from the far-infrared (FIR) to UV luminosity ratio. Here we used the A$_{\rm IRX}$\,$-$\,$\beta$ relationship derived by \citet{talia15}. All these measurements are presented in Table~\ref{sfr}.

The overdensity members $m1-m6$ have generally blue spectra 
($\beta \sim-1.3$ to $-2.7$) and are actively forming stars, with extinction-corrected SFRs in the range $\approx 8-60 \, M_{\odot}$/yr (see Table~\ref{sfr}). For the serendipitous LUCI source $l2,$ we derived
a star formation rate of $>$5 $M_{\odot}$/yr based on its H$\alpha$ luminosity (e.g., \citealt{kennicutt12}). We consider the SFR measured for $l2$ as a lower limit because we did not apply any correction for the extinction. We did not find any evidence in our LBT/LBC and CFHT/WIRCam images for objects with optical or near-IR colors that would be consistent with a sequence of red passive galaxies at $z\sim1.6-1.7$ ($i-J>1.5$, \citealt{chan18}), except for the optically faint ($i_{AB}\sim26$) radio object shown in Fig.~\ref{sepia} (for which we measured a photometric redshift consistent with 1.7, see Section \ref{arinp}). 
This again suggests that the structure is young and likely not yet virialized, as often found for protoclusters around HzRGs \citep{overzier05,kotyla16}. 

We used the STARLIGHT full-spectrum fitting code \citep{cid05} to recover the star formation history of galaxies $m1-m6$. STARLIGHT fits an observed spectrum with a superposition of synthetic simple stellar populations (SSPs) of various ages and metallicities, producing a best-fit spectrum. In particular, it provides the light and mass contribution of each synthetic SSP to the best-fit model at a user-defined normalization wavelength (set  to 2500 $\AA$ in our fits). This allowed us to recover the galaxy star formation history as the distribution of the SSP contributions as a function of their age. We adopted Bruzual $\&$ Charlot (2003) SSP models (updated to 2016 - BC16 hereafter), which extend down to 1000 $\AA$ with a resolution $\rm \Delta\lambda=1~\AA~FWHM$. We downgraded the spectral resolution of the BC16 models to that of the observed spectra, which is $\rm \lambda/R/(1+z) \sim$ 2.5 $\AA$, considering $R~\sim~1000$, $z=1.7$ and an average observed $\rm \lambda~\sim$ 6500 $\AA$. We adopted a grid of metallicities that reached from 0.005 $\rm Z_{\odot}$ to 5 $\rm Z_{\odot}$ and ages from 100 Myr to 4 Gyr, which corresponds to the age of the Universe at $\rm z=1.7$. 
Because the synthetic SSPs only model the stellar contributions to the UV flux and do not account for nebular emission or absorption from the interstellar medium (ISM), we performed the spectral fitting by masking the strongest nebular emission lines (i.e., $\rm C III]~ \lambda\lambda1906.68,1908.68$ and $\rm C II]~\lambda2326.00$) and absorption lines including ISM absorption (i.e., $\rm Al~III $ and $\rm Fe~II $ absorption lines in the range 1850 - 2850 $\AA$, see Table 1 in \citealt{talia12}). 
Significant episodes of star formation as young as a few megayears are found in all galaxies. Furthermore, in $m1$ and in $m4$ these recent bursts produce most of the optical and UV light. 

\begin{figure}[t]
\includegraphics[angle=0, width=9cm]{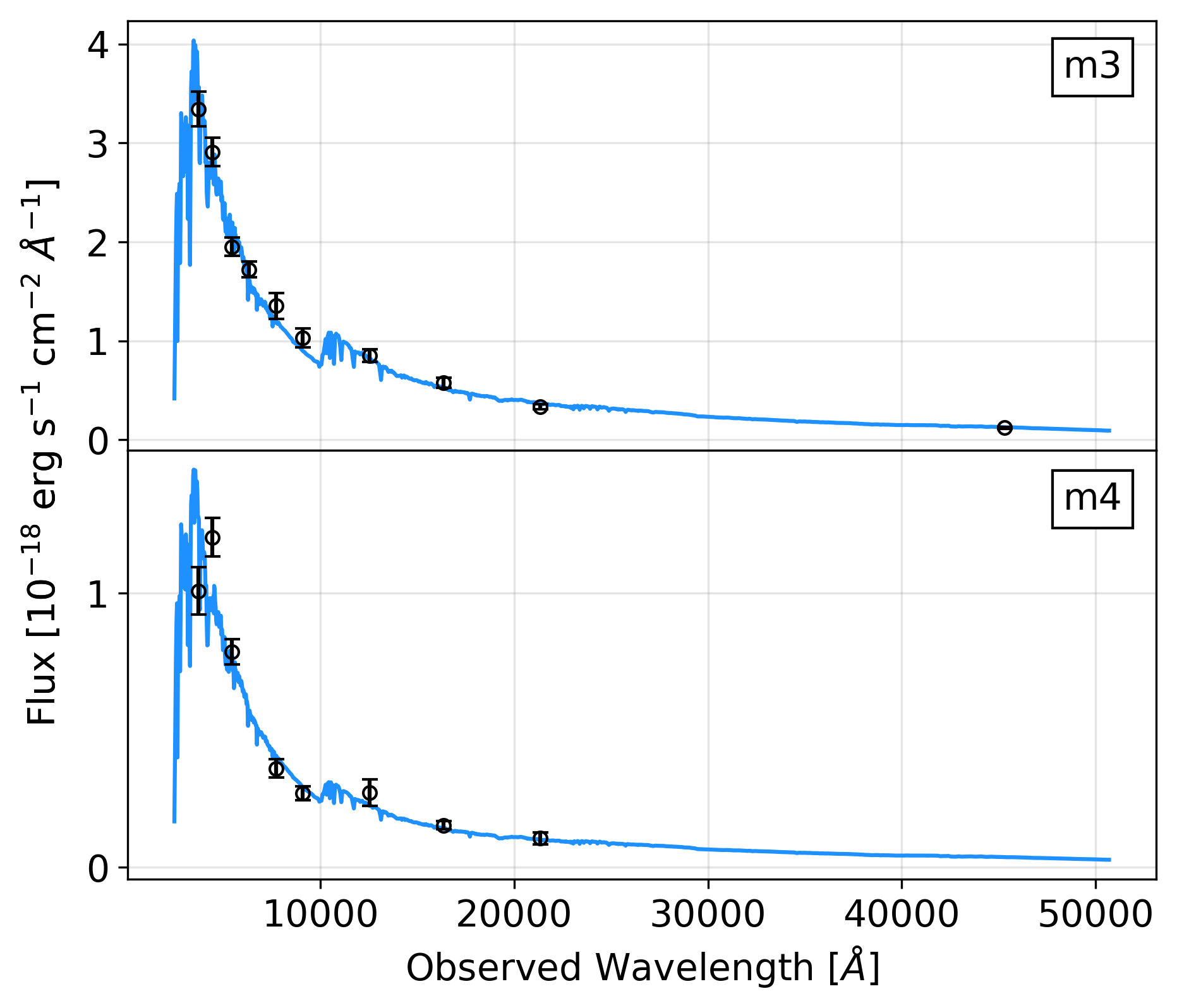}
\caption{Spectral energy distribution of the overdensity galaxies $m3$ and $m4$ and best-fit model.}
\label{sedfit}
\end{figure}

We derived the stellar masses $M_*$ of $m1-m6$ by means of an SED fit using $Hyperzmass$, a modified version of the $Hyperz$ code (see, e.g., \citealt{bolzonella10}).
The fit is based on Bruzual $\&$ Charlot (2003) SSPs assuming different star formation histories (either an exponentially declining or a constant SFR).
Reddening was also introduced following Calzetti et al. (2000). We collected all photometric data points available for $m1-m6$, from the U-band to IRAC
3.6 and 4.5$\mu$m. From 6 to 12 photometric detections at different wavelengths are available for our sources.  
We verified that in the MUSYC catalog the photometry of a few sources
is highly inaccurate because of the contamination from the bright foreground star, in particular in the i and z bands. When available, we then used HST photometry
in the F775W and F850LP filters (together with the F160W filter)\footnote{Based on the Hubble Source Catalog, see: https://archive.stsci.edu/hst/hsc/}. The stellar masses obtained for $m1-m6$ are reported in Table~3. Two examples of
an SED fit (for $m3$ and $m4$) are shown in Fig.~\ref{sedfit}. The best-fit SFRs obtained with the SED fitting agree within the errors with those derived
from the UV spectra. The only discrepancy is found for $m6$, for which the SED fit derived a ten times lower SFR than what was estimated from the fit to the 
UV spectrum, where a relatively flat continuum was interpreted as a highly extincted starburst. The SED fit instead reveals a strong Balmer break outside the wavelength range 
covered by the UV spectrum, suggesting that the red spectral color is instead due to evolved stellar populations. For $m6$ we then report the
SFR measured with $Hyperzmass$ rather than from the UV spectral fit. In Table 3 we also provide the specific star formation rates (sSFR=SFR/$M_*$) 
obtained by combining the results of the UV spectral and SED fit.

\begin{table}[t]
\begin{center}
\begin{tabular}{cccccr}
\hline \hline
ID& $M_{2800}$& $\beta$& SFR& log$(M_*)$& log(sSFR)\\
& (AB)& & ($M_{\odot}\;yr^{-1}$)& ($M_{\odot}$)& $(Gyr^{-1})$\\
(1)&(2)&(3)&(4)&(5)&(6)\\
\hline
$m1$& -19.52& -1.33& 20.2$\pm3.4$& 9.36$^{+0.47}_{-0.64}$& 0.94$\pm$0.55\\
$m2$& -19.53& -1.45& 18.1$\pm3.4$& 9.69$^{+0.09}_{-0.04}$& 0.57$\pm$0.11\\
$m3$& -21.25& -1.88& 57.1$\pm6.2$& 10.09$^{+0.05}_{-0.07}$& 0.67$\pm$0.08\\
$m4$& -20.08& -2.22& 13.8$\pm1.1$& 9.33$^{+0.19}_{-0.28}$& 0.81$\pm$0.23\\
$m5$& -19.96& -2.70&  7.6$\pm1.1$& 9.54$^{+0.16}_{-0.36}$& 0.34$\pm$0.28\\
$m6$& -19.94& -0.35&  13.1$\pm1.0$$^a$& 10.48$^{+0.04}_{-0.01}$& -0.37$\pm$0.04\\
\hline                                             
\end{tabular}
\caption{Properties of MUSE members of the overdensity. Columns: 1) source identifier; 2) absolute magnitude at 2800\AA\ rest-frame (not corrected for extinction); 3) slope of the optical spectrum ($f_\lambda\propto \lambda^\beta$); 4) extinction-corrected star formation rate based on the UV spectrum; 5) stellar mass; and 6) specific star formation rate.
Notes: $^a$Based on the SED fitting because it provides a more reliable estimate for this source (see text for details). 
} 
\label{sfr}
\end{center}
\end{table}

\subsection{Power, orientation, and nuclear obscuration of the FRII}\label{power}

Our Chandra spectral analysis indicates that the AGN powering the FRII radio-galaxy is a heavily obscured Compton-thick QSO (XID189). The obscuration measured in the X-rays is 
consistent with the type 2 optical classification derived from the LUCI spectrum. According to the classic unification schemes, the highest 
column densities are expected in systems where the inclination angle $\theta$ between the axis of a toroidal $\text{approximately }$parsec-scale distribution of obscuring gas and our line of sight is $\sim 90$ deg, 
that is, when the system is seen edge-on. In the hypothesis that the obscuring torus and radio jet are coaxial, we can estimate $\theta$ based on the radio data. We used the VLA radio images to estimate the flux ratio $R_{jet}$ between the jet versus counter-jet emission at the smallest possible scales and obtained a joint constraint on the jet velocity $\beta=v/c$ and inclination angle $\theta$ (assuming equal power jets) of
\begin{equation}
R_{jet} = [(1+k)/(1-k)]^{p+\alpha_{jet}}, 
\label{rjet}
\end{equation}

where $k \equiv \beta cos(\theta)$,  $\alpha_{jet}$ is the spectral index of the jet ($f_{\nu}\propto \nu^{-\alpha_{jet}}$) and $p$ is the Doppler boost exponent. 
We measured $R_{jet}$ in the range $1.42$-1.65, which for typical values of $\alpha_{jet} \sim 0.5$ and assuming a continuous jet with $p=2$ translates into
$k\sim 0.07-0.1$. 

Another constraint on $k$ can be obtained by comparing the observed Doppler-boosted power in the core  $P_{core}^{obs}$ with an estimate of the intrinsic
(not boosted) core power  $P_{core}^{int}$ (see, e.g., \citealt{cohen07}):
\begin{equation}
R_{core}\equiv P_{core}^{obs}/P_{core}^{int}=\delta^{p+\alpha_{core}},
\label{rcore}
\end{equation}

where  $\delta = \sqrt{(1-\beta^2)}/(1-k)$ is the relativistic Doppler factor. By combining Eqs.\ref{rjet} and \ref{rcore}, we can then derive both $\beta$ and $k$, and in turn, 
the inclination angle $\theta=arcos(k/\beta)$. 

An estimate of the intrinsic core power $P_{core}^{int}$ can be obtained from the total extended radio power by means of the $P_{tot}(408MHz) - P_{core}^{int}(5GHz)$ correlation described, for instance, by \citet{giovannini01}.
Based on the 150 MHz data of the GMRT TGSS survey\footnote{\url{http://tgssadr.strw.leidenuniv.nl/doku.php}}, we measured a total flux in the
two summed radio lobes equal to $\sim160$ mJy (within the $3.5\sigma$ contours). For a radio spectral index of $\alpha_{tot}=0.8\; (f_{\nu}\propto \nu^{-\alpha_{tot}})$ that we derived from the comparison between the TGSS data and the NVSS data at 1.4 GHz (see \citealt{nanni18} for details), the TGSS flux converts into a total rest-frame extended luminosity at \
408MHz of $P_{tot}(408MHz)\sim 10^{26}$ W~Hz$^{-1}$sr$^{-1}$. This value, when combined with the $H\alpha$ luminosity measured in Section~\ref{luci}, nicely places XID189 on the
total radio power versus $H\alpha$ luminosity correlation reported by \citet{zirbel95} for powerful FRII radio galaxies.

The observed core power $P_{core}^{obs}$ at 5GHz was assumed to be equal to that observed at 1.4GHz (i.e., we assumed a typical spectral slope of $\alpha_{core}=0.0$ because this cannot be
derived directly from our data), which gives $R_{core}\equiv P_{core}^{obs}/P_{core}^{int}\sim 0.9-1.2$. 
Based on Eqs.\ref{rjet} and \ref{rcore}, we finally derived $\beta \sim 0.4-0.5$ and $\theta \sim 70-80$ deg. This indicates that the system is truly seen almost edge-on, as expected for this heavily obscured FRII nucleus.

The absorption-corrected luminosity in the rest-frame 2-10 keV band is $L_{2-10}=1.3\times 10^{44}$\ergs, which, adopting a bolometric correction of 30, as appropriate 
for these X-ray luminosities (e.g., \citealt{marconi04}), translates into a total radiated luminosity of $L_{rad}\sim4\times 10^{45}$ erg s$^{-1}$.\ This is well into the QSO regime. 
To compute the bulk kinetic power of the FRII jet, we assumed it to be proportional to the total radio luminosity of the lobes, following \citet{willott99} (see also \citealt{hardcastle07} and \citealt{shankar08}):
\begin{equation}
P_{jet} = 3\times 10^{45}  f^{3/2}  L_{151}^{6/7}   erg s^{-1},
\label{pjet}
\end{equation}

where $L_{151}$ is the total observed luminosity at 151MHz in units of $10^{28}$ W~Hz$^{-1}$sr$^{-1}$ and the factor $f$ encapsulates all the systematic uncertainties on the system geometry and environment and on the jet composition (see \citealt{willott99} for details). 
Based on the TGSS data and on the spectral slope mentioned above, we derived a rest-frame extended radio luminosity of $L_{151} \sim 2.1\times 10^{26}$  W~Hz$^{-1}$sr$^{-1}$. 
By further assuming $f=15$, which is a reasonable value for FRII radio galaxies (see \citealt{hardcastle07}), from Eq.~\ref{pjet} we derived a total jet kinetic power of  $P_{jet} \sim 6.3\times 10^{45}$ erg~s$^{-1}$. Our measurements indicate that $P_{jet}\sim 1.5\times L_{rad}$, that is, the energy released in the form of baryons in the jet is equal to or even higher than that
released in the form of photons by the accretion disk,  as is generally found for AGN with powerful radio jets \citep{ghisellini14}. The effects of this energy release on the surrounding environment is discussed in Section \ref{feedback}.

\begin{figure}[t]
\includegraphics[angle=0, width=9cm]{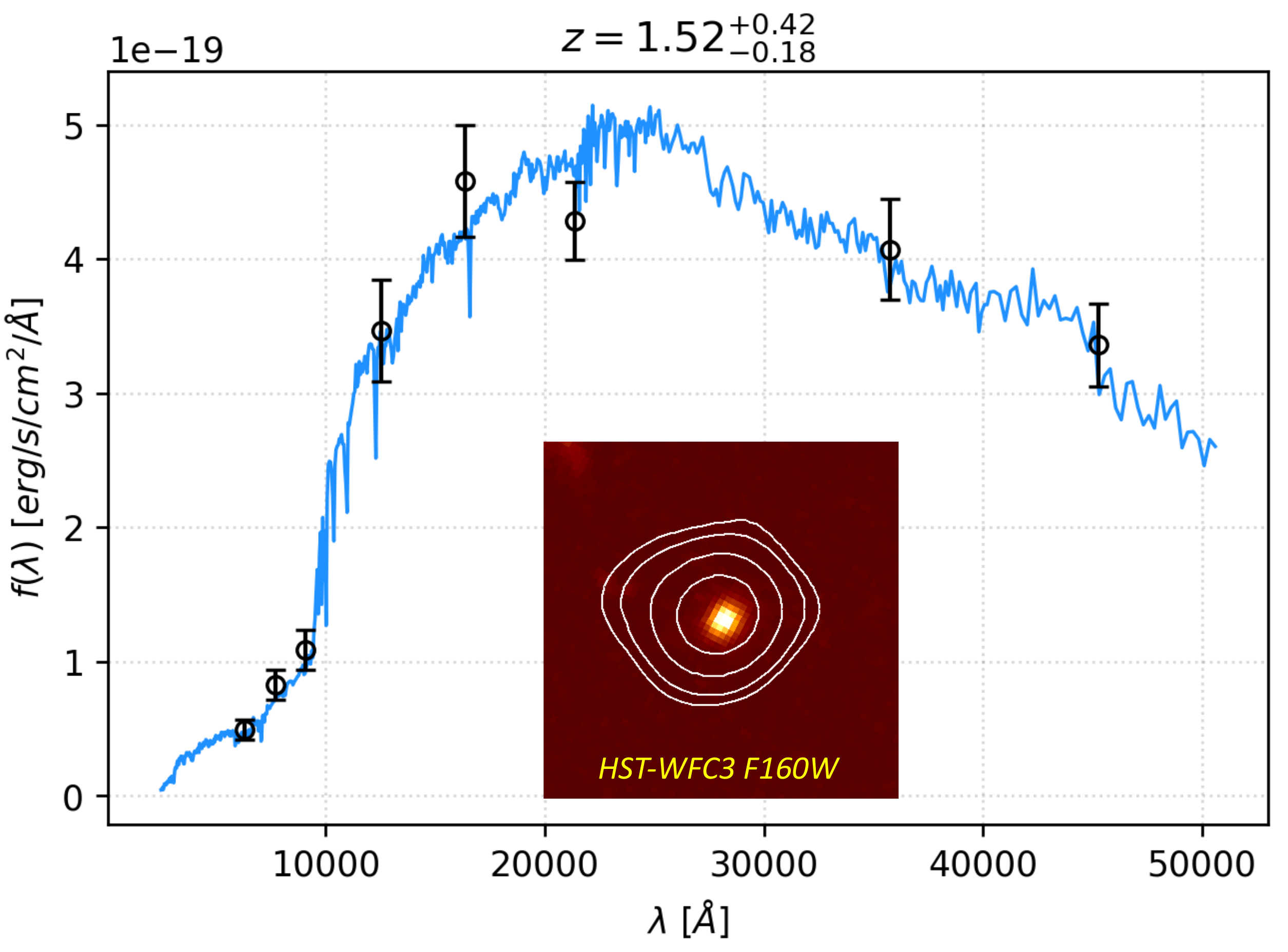}
\caption{Broadband SED and photometric redshift solution for the radio source at the edge of component A of the diffuse X-ray emission. The inset shows an 8"x8" cutout of the HST/WFC3 F160W image with radio contours overlaid in white (with the same levels as used in Fig.~\ref{xid189}).}
\label{zphot}
\end{figure}

\begin{figure*}[t]
\begin{center}
\includegraphics[angle=0, width=15cm]{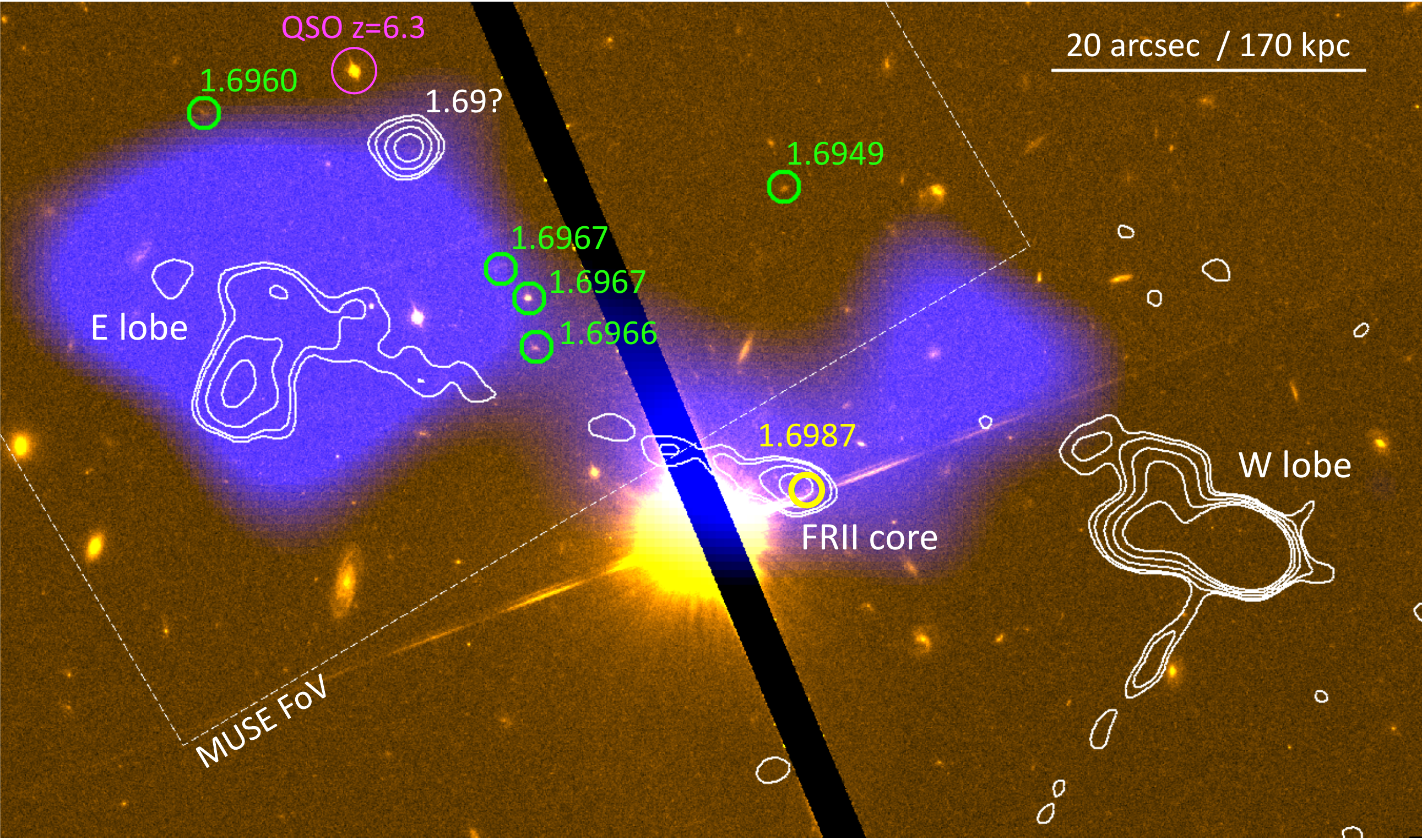}
\end{center}
\caption{HST/ACS F850LP image of the overdensity overlaid with radio contours from the VLA (in white, same levels as in Fig.~\ref{xid189}) and the Chandra/ACIS-I smoothed image of diffuse X-ray emission (blue and violet; 0.5-7 keV band, point sources  removed). North is up and east is to the left. The dark strip running across the bright star is the gap between the two ACS CCDs. The main radio morphological features of the FRII galaxy are labeled in white. The position of the MUSE pointing is shown. The position of the FRII host and its redshift are shown in yellow. Green circles mark MUSE galaxies in the overdensity; their redshifts are as labeled. The additional radio source that may be part of the overdensity ($1.4<z_{phot}<1.9$ at 68\% confidence level) is labeled in white. The position of the z=6.3 QSO SDSS~J1030+0524 is also marked in magenta.}
\label{sepia}
\end{figure*}

\section{Discussion}

\subsection{Total mass of the structure}\label{mass}

Measuring the total mass of a system that is likely far from virialization is not an easy task. A rough estimate of the dark matter mass contained in the structure can be obtained by means of the measured galaxy overdensity: $M_{tot}= \overline\rho_m V (1+\delta_m)$, where $\overline\rho_m$ is the average density of the Universe at the redshift of the structure, V is its volume, and $\delta_m$ is the dark matter overdensity. In order to have some control on the volume spanned by our observations, we considered in our estimate only the six galaxies found by MUSE.
The volume $V$ is assumed to be a box of dimensions $0.5 \times 0.5 \times 6.3$ = 1.6 proper Mpc$^3$, that is, the MUSE FoV ($0.5 \times 0.5$ Mpc$^2$ at $z=1.7$) multiplied by the maximum radial separation of the MUSE overdensity
members (i.e., 6.3 Mpc assuming that peculiar velocities are negligible; see, e.g., Fig.~\ref{dv}). The dark matter overdensity can be derived as $\delta_m = \delta_g/b$, where $\delta_g$ is the measured galaxy overdensity ($\delta_g=22$, see Section~4.1), 
and $b$ is the bias of the considered galaxy population. Here we  assumed $b=2$, which is appropriate for galaxies with star formation rates and redshifts similar to those of our MUSE sample (e.g., \citealt{lin12}). Under these assumptions the derived structure mass is $M=1.3\times 10^{13}\; M_{\odot}$. 
If the system is far from virialization and still collapsing, galaxy infall motions would cause the system volume $V_{app}$ to appear smaller and the overdensity in turn higher than their true values.
We tried to correct for these effects and estimated the volume compression factor $C$ following \citet{cucciati14}. The true volume of the system
is defined as $V_{true}=V_{app}/C$ (with $C<1$) and the true mass overdensity $\delta_m$ is given by $1 + b\delta_m = C(1+\delta_{g,obs})$, where 
$\delta_{g,obs}$ is the observed galaxy overdensity, and $b=\delta_g/\delta_m$ is the true bias parameter. In the hypothesis that the system is undergoing a simple spherical collapse, the compression factor $C$ can be written as
$C = 1 + f - f(1+\delta_m)^{1/3}$  \citep{steidel98}, where $f(z) \approx \Omega_m(z)^{0.6}$ for a $\Lambda$CDM model \citep{lahav91}. By assuming again $b=2$ and by solving numerically, we obtained $C$=0.4, that is, $V_{true} = 3.9$~Mpc$^3$, and $\delta_m$=3.7. Using these revised values, we obtained for the system a mass of $M=1.5\times 10^{13}\; M_{\odot}$, very similar to what was 
estimated by the observed values, as the larger volume is largely compensated for by the lower overdensity. 
By considering that the structure likely extends far beyond the MUSE FoV (see, e.g., Fig.~\ref{full}), these measurements are presumably lower limits. For comparison, the total virial mass that would be obtained for this system from the observed line-of-sight velocity dispersion of 325 \kms(following, e.g., \citealt{lemaux12} and \citealt{cucciati18}) would be $\approx 2.5\times 10^{13}$\msun. Future spectroscopic observations on wider areas are needed to properly sample the size and mass of the structure and verify 
whether this is the progenitor of a massive galaxy cluster ($M>10^{14}\;M_{\odot}$) caught in its major assembly phase.

\subsection{Another radio galaxy in the structure?}\label{arinp}

The VLA data showed the existence of another radio source within the region covered by the overdensity. As shown in Figs.~\ref{csmooth} and \ref{HR}, this additional radio source is compact and falls at the northern edge of component A of the diffuse X-ray emission, in between MUSE galaxies $m1$ and $m2$. It is therefore interesting to investigate whether it is part of the structure. The source is faint in the optical images 
(e.g., $r=26.18\pm0.13$, $i=25.35\pm0.13$ in the LBT/LBC catalog of \citealt{morselli14}) and has very red colors (e.g., $i-J=2.5$, using LBT/LBC and CFHT/WIRCam photometry). 
We computed the photometric redshift of the radio source by running the $Hyperz$ code \citep{bolzonella00} on the broadband
photometry obtained from the LBT/LBC, WIRCam, and MUSYC catalogs. We also added IRAC photometry at 3.6 and 4.5 $\mu$m as retrieved from the Spitzer public catalog available at IRSA\footnote{https://irsa.ipac.caltech.edu/}. The broadband photometry and associated best-fit redshift solution are shown in Fig.~\ref{zphot}. The measured photometric redshift $z_{phot}=1.52^{+0.42}_{-0.18}$ (1$\sigma$ errors) is fully consistent with $z=1.69$, and the radio source is therefore an additional candidate AGN in the structure that is hosted by a passive evolved galaxy.
Assuming a redshift of z=1.69, the total flux density of $300 \pm 30\; \mu$Jy measured in the VLA images at 1.4GHz converts into a total radio power of $L_{1.4GHz}=4.7\pm0.5 \times 10^{23}$ \whz sr$^{-1}$, placing this source just above the knee of the radio luminosity function of radio-emitting AGN at comparable redshifts \citep{smolcic17}. Moreover, based on the SED fit with $Hyperzmass$ and using z=1.69, we measured a stellar mass for this object
of $M_*\sim 10^{11}\;M_{\odot}$. This value nicely agrees with what we derived using the tight correlation between stellar mass and observed K-band magnitude found by \citet{nantais13} for the members of two galaxy clusters at z=0.8 and z=1.2 (rms of 0.14-0.19 dex), and rescaling to z=1.7. For comparison, the stellar mass that we obtained for the FRII host, that is, for the candidate brightest cluster galaxy (BCG) progenitor using its K-band magnitude, is only three times higher. Because of its vicinity
to the bright central star, no reliable photometry can be obtained in bluer bands for the FRII host, which prevents any accurate SED fitting. If future optical and near-infrared (NIR) spectroscopy confirms that the radio source is at $z_{spec}$=1.7, then it will be one of the most massive galaxies of the overdensity.

\subsection{IC-CMB as the origin of the diffuse X-rays?}\label{diffx}

We here investigate the possibility that the diffuse X-ray emission seen in component A arises from inverse Compton scattering of CMB photons by the relativistic electrons of the eastern radio lobe (IC-CMB), as the energy density of the CMB steeply increases with redshift as $(1+z)^4$. The hard spectrum of component A (see Table~\ref{xtab}) might indeed indicate a non-negligible contribution from IC-CMB. As discussed in \citet{nanni18}, the absence of X-rays from the western lobe, which is a factor of $>6$ times brighter in the radio, poses a challenge to this interpretation. We investigate this question in more detail below.

If the energy in the radio lobes is equally distributed between relativistic particles and magnetic field, it is possible to provide an estimate of the magnetic field at the equipartition $B_{eq}$, and in turn, of the flux density at 1 keV expected from IC-CMB using, for instance, Equation (11) of \citet{harris79}:
\begin{equation}
f_{1 keV}=\frac{(5.05 \times 10^4)^{\alpha} C(\alpha) G(\alpha)(1+ z)^{\alpha+3} f_{r}\nu_{r}}{10^{47}B_{eq}^{\alpha+1}},
\end{equation}

where C($\alpha$) and G($\alpha$) are tabulated constants (see, e.g., \citealt{harris79} and \citealt{pacholczyk70}) and $\alpha$  is the radio spectral index ($f_r \propto \nu_r^{-\alpha}$).
The flux densities, measured in the same region, are in cgs units.

$B_{eq}$ can be calculated using Eq. (3) of \citet{miley80} and following standard prescriptions, that is, a ratio of energy in the heavy particles to that in the electrons k=1, a volume filling factor $\eta= 1$, and an angle between the uniform magnetic field and the line of sight $\varphi=90^{\circ}$ .
For the eastern lobe we considered a radio spectral index  of $\alpha=0.8$ (0.01 GHz $<\nu < 100$ GHz) and a flux density of 
$f_{r}^{obs}$=1.7 mJy at $\nu_{r}^{obs}=1.4$ GHz as derived from the VLA maps. 
The emitting region (E lobe in Fig.~\ref{sepia}) was approximated by an ellipse of angular diameters $\theta_x \sim 10$ arcsec, $\theta_y\sim 12$ arcsec. The path length through the source in the radial direction was assumed equal to the angular diameter in the x-direction, that is,  $\sim 90$ kpc.
We estimated a magnetic field of $B_{eq}\sim 5$ $\mu$G, in reasonable agreement with values reported in the literature \citep{isobe11}.
Under this condition, the IC-CMB flux in the 0.5-7 keV band (again assuming a power law of spectral index $\alpha$=0.8, i.e., a photon index of $\Gamma=1.8$, consistent within the errors with that reported in Table~2 for component A) is 60 times lower than what is observed in component A.
Even considering a population of relativistic electrons missed by current VLA data and distributed over the entire region A, the expected X-ray flux would still be more than one dex lower than what is observed. It is nonetheless possible that the extended radio structures are not in an equipartition state, as observed in some galaxies (\citealt{migliori07,isobe11}). In this case, a reduction of the magnetic field by a factor of $\sim 3$ would be sufficient to obtain the observed X-ray flux in the A region. 
If such a deviation from equipartition is present in both radio lobes, then an X-ray flux of $\sim 10^{-15}$\cgs has to be expected in the western lobe, which is instead excluded by our deep Chandra data (at the $\sim 2\sigma$ level). Therefore, to simultaneously explain the diffuse X-ray emission around the eastern lobe and its absence around the western lobe,
we have to assume that 1) the magnetic field in the eastern lobe is a factor of 3 below equipartition, 2), the western lobe is instead around equipartition, and 3) low surface brightness emission that fills the entire component A has been missed by current radio data. We consider such a combination of requirements less probable than the shock-heating scenario discussed in the next section, but future observations in the radio band, for example, with the Low Frequency Array (LOFAR), will reveal whether IC-CMB is a plausible scenario as well.

\subsection{Star formation promoted by AGN feedback}\label{feedback}

Intriguingly, the overdensity members $m1-m4$ (as well as the radio object at $z_{phot}\approx1.7$) appear to lie at the northern boundary of component A of the diffuse X-ray emission, that is, the one around the eastern radio lobe of the FRII (see Figs.~\ref{csmooth} $left$ and ~\ref{sepia}). We tested whether the location of the boundary of component A, and hence its spatial coincidence with the position of the MUSE sources in the overdensity, is a spurious effect introduced by our smoothing procedure or if it is a real feature. We computed the radial profile of the X-ray surface brightness of component A on the unsmoothed 0.5-7 keV image using the series of annuli shown 
in Fig.~\ref{csmooth} ($left$). We evaluated the background in a nearby region free from point-like or diffuse X-ray sources. The surface brightness profile, shown in Fig.~\ref{csmooth} ($right$), rapidly falls off beyond 
$\sim14$ arcsec from the center of component A, and it is consistent with zero beyond that distance. Similar results are obtained for the 0.5-2 keV image. The four overdensity members $m1-m4$ are all at this roll-off distance (see Fig.~\ref{csmooth} and~\ref{sepia}). We therefore conclude that the spatial coincidence between the boundary of the diffuse X-ray emission and the location of the four galaxies is not an artifact introduced by our smoothing procedure. Statistically, the binomial probability of finding
four (or more) out of the six z=1.69 MUSE galaxies along that boundary (assuming it can be approximated with an annulus of radius 14" and width 5", hence representing $\sim$5\% of the MUSE FoV)  is very low, $P\sim1.2\times 10^{-4}$. 

The spatial overlap between the radio lobe and the diffuse X-ray emission of component A and the location of the $m1-m4$ members
of the overdensity, right at the edge of this diffuse X-ray emission, led us to speculate that there is a physical connection between all these observables. For
example, the eastern lobe of the radio galaxy bends toward the south, indicating that some interaction
with hot gas within component A is plausible.

We first investigated whether, assuming that the observed X-ray emission is mostly thermal, the jet is powerful enough to deposit the observed amount of energy into component A. Based on its measured projected size of 240 kpc, we approximated component A as a sphere of 120 kpc radius. By assuming a gas temperature and density of $T=$5 keV and $n\sim 4\times10^{-3}$ cm$^{-3}$, respectively (consistently with the spectral fit), we find that a total thermal energy of $E_{th}\sim nVkT\sim 7\times 10^{60}$ erg is stored in component A. For comparison, the absence of X-ray emission in the western lobe ($f_{0.5-7}< 1.4\times 10^{-15}$\cgs at $2\sigma$) translates into a 2$\sigma$
upper limit on the gas density of $n<2.5\times10^{-3}$ cm$^{-3}$ (assuming that this gas has the same extension and temperature as the gas measured in the eastern lobe, and hence that the difference in the X-ray emissivity only depends on  $n^2$). This limit would be even less stringent 
if the putative hot gas associated with the W lobe is not as extended as component A. We therefore conclude that the observed asymmetry in the X-ray morphology
may be caused by relatively small variations in the particle density of the shock-heated gas.

Both recent hydrodynamic simulations of AGN jet propagation within intracluster medium \citep{bourne19} and classic computations
of energy-driven outflows produced by steady winds in astrophysical sources \citep{w77} suggest that about 40-50\% of the jet or wind power goes into gas heating,
whereas the other half goes into $pdV$ work on the ambient medium. By assuming a constant jet power equal to that measured in Section~\ref{power}, 
$P_{jet} = 6.3\times 10^{45}$ erg~s$^{-1}$, and that only half of it is available for gas heating, it would take
$E_{th}/(P_{jet}/2)=70$ Myr to heat the gas up to the level observed in component A. This lifetime is perfectly consistent with the spectral ages of $\sim70-80$ Myr recently derived from LOFAR observations at low radio frequencies of local FRII radio galaxies of similar power to XID189 \citep{harwood17}. We recall that because of the uncertainties in the actual gas temperature, 
the total thermal energy of component A is also uncertain. On the one hand, if $T>5$ keV (as derived from the X-ray spectral fit when a pure thermal model is used), the total thermal energy would be higher. On the other hand, because of the likely contribution of IC-CMB emission to the observed X-ray luminosity, the temperature and thermal energy of the gas in component A may have been overestimated, and the constraints on the duration of the jet activity can in turn be relaxed. These arguments suggest that the thermal energy reservoir of component A may plausibly be produced by AGN feedback. 

\begin{figure}[t]
\includegraphics[angle=0, width=9.cm]{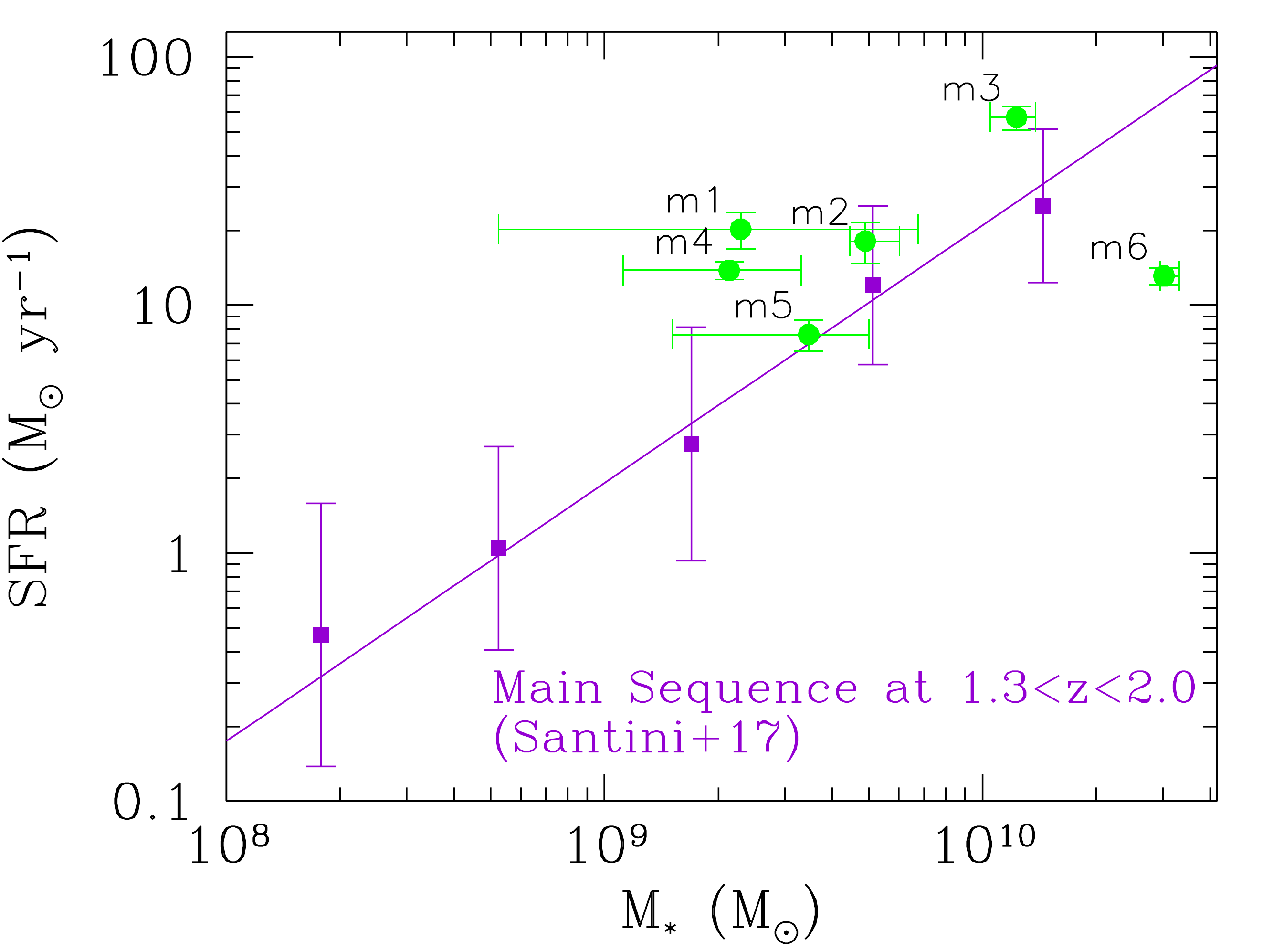}
\includegraphics[angle=0, width=9.cm]{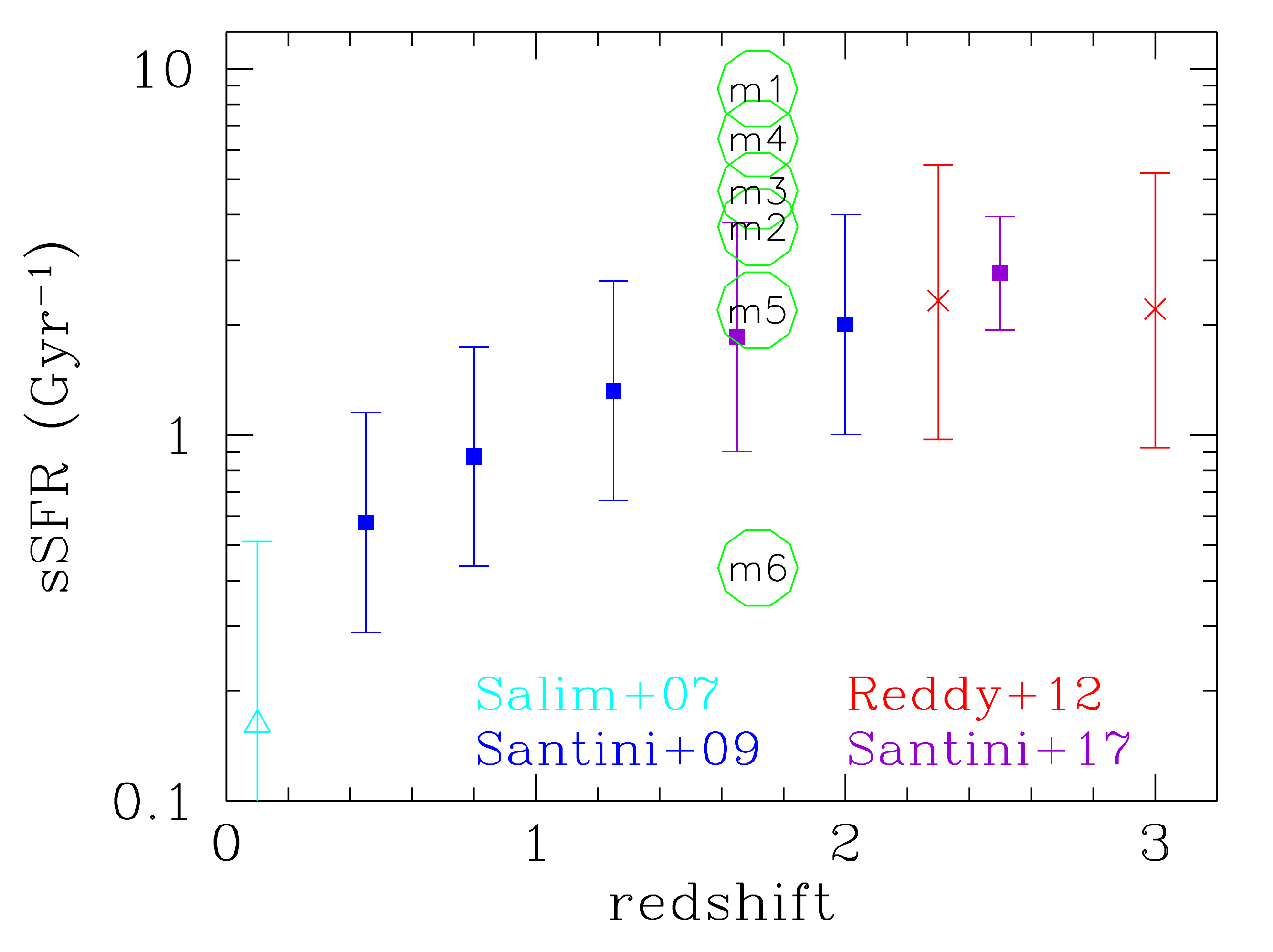}
\caption{{\it Top panel:} Star formation rate vs. stellar mass for the overdensity galaxies $m1-m6$ (green circles). The violet squares and line show the main sequence of field galaxies at $1.3<z<2.0$ and its best-fit relation as derived by \citet{santini17}. {\it Bottom panel:}  Specific star formation rate of the galaxies $m1-m6$ (green circles) plotted against the specific star formation rate vs. redshift relation derived for field galaxies with log$(M_*/M_{\odot})=9.5-10.0$. Literature data points are labeled.}
\label{ssfr}
\end{figure}

Furthermore, if the X-ray emission of component A is due to a bubble of gas that is shock-heated by the FRII jet, we may wonder whether this bubble is still expanding or whether it has stalled at the observed boundary. We compared the internal gas pressure of component A ($P_{hot}$) with that of a putative cold ambient medium within the overdensity ($P_{cold}$), 
assuming this shares the same physical properties of the CGM observed in the Spiderweb protocluster, where a reservoir of $10^{11}\; M_{\odot}$ of cold molecular gas has been found to extend on scales of $\sim50-70$ kpc diameter through ATCA and ALMA observations of low-J CO transitions \citep{emonts18}. Based on these measurements, the particle density of the ambient medium would be $n\sim 7\times 10^{-2}$ cm$^{-3}$, and its temperature is expected to be a few tens of $\text{kelvin}$, which would imply $P_{hot}/P_{cold}>10^5$, that is,
the hot gas bubble is still expanding. A similar result ($P_{hot}/P_{cold}\gtrsim10^4$) is obtained if instead the ambient medium is mostly in the form of atomic gas with temperatures $T<10^5 \; K$ and column densities $N_H\sim 10^{19-20}$ \cm, similar to those found in protoclusters at $z\sim2$ through the study of absorption lines (e.g., Ly$\alpha$, Si II, and CIV) in the rest-UV spectra of background objects \citep{cucciati14,hennawi15}. Again, with the caveat that we might have somewhat overestimated the pressure of the hot component, the bubble of hot gas is likely expanding with time.

We remark that to our knowledge, this is the first time that such a hot gas bubble inflated by AGN feedback is observed in a high-z structure that is likely caught in its
main assembly phase. Such a hot bubble may contribute significantly to the pre-heating of the ICM by mixing with the cooler ambient gas (see, e.g., \citealt{hillel16}). The rapid expansion of this bubble (v $>$1000 \kms, see below) and the consequent ICM preheating on scales of up to several hundred kiloparsecs may even be favored in the early evolutionary
stages of cluster assembly, when the pressure and density of the ambient gas are far lower than those of evolved and collapsed structures. All these possibilities, as well
as the observational prospects of detecting these early hot gas bubbles, for instance, through the Sunyaev-Zeldovich effect (see, e.g., \citealt{ehlert19}), will be discussed elsewhere. 

It is now interesting to verify whether the dimensions of the hot gas bubble are consistent with its expansion velocity and age, that is, with the AGN lifetime of 70 Myr derived above. At $v=0.5c$, it would only take 2 Myr for the jet particles to move from the FRII core to the eastern lobe, therefore the bubble age can be considered equal to the AGN lifetime. Several works address the problem of expanding gas bubbles inflated by shocks produced by relativistic collimated jets or relativistic
wide-angle winds launched by an AGN, using both analytic and numerical methods \citep{w77,costa14,bourne19}. We used here the formalism presented in \citet{gilli17}, where we derived the expansion laws for bubbles that are inflated by relativistic winds launched by exponentially growing supermassive black holes. For simplicity, we assumed that a shock is generated at the center of component A as the relativistic jet collides with a clump of cold gas in the structure, and that this shock expands isotropically. As the cooling time of the shock-heated gas is expected to be longer than the AGN lifetime, we refer to the case of an energy-driven
outflow, that is, to Eqs. 16 and 15 in \citet{gilli17}, which describe the temporal evolution of the bubble radius $R(t)$ and its normalization $R_0$, respectively\footnote{Note that $R_0$ {\it is not} $R(t=0)$; see \citet{gilli17}.}. These equations, which were derived for radio-quiet AGN, can be extended to jetted sources
by considering that the gravitational energy of the matter falling toward the black hole is dissipated not only through radiation from the accretion disk ($L_{rad}$), but also through jet launching ($P_{jet}$). The global accretion efficiency $\epsilon_{tot}$ can therefore be written as $\epsilon_{tot} = \epsilon_r + \epsilon_k$, 
where $\epsilon_r$ and $\epsilon_k$ are the disk radiative and the jet kinetic efficiency, respectively (see, e.g., \citealt{jolley09} and \citealt{ghisellini13}).
Accordingly, the black hole mass grows from its initial value $M_0$ as $M(t)=M_0 e^{t/t_{Sal}}$, where the Salpeter time $t_{Sal}$ now reads:
$t_{Sal}=0.45\; Gyr \;\left(\frac{\epsilon_r}{1-\epsilon_{tot}}\right)\lambda^{-1}$, where $\lambda$ is the Eddington ratio.
The normalization of the bubble radius $R_0$ depends on several parameters: the Salpeter time $t_{Sal}$, the ratio $f_w\equiv P_{kin}/L_{rad}$, the Eddington ratio $\lambda$, the initial black hole mass $M_0$, and the ambient gas density $\rho_{gas}$. We assumed here that the black hole is accreting at its Eddington limit ($\lambda=1$), and that it is rapidly spinning (with, e.g., $\epsilon_{tot}=0.3$), as is generally thought for black holes powering jetted AGN \citep{blandford77,ghisellini14}. 
The $L_{rad}$ and $P_{kin}$ values measured in the previous sections give $f_w\sim 2$ (i.e., $\epsilon_r \sim 0.1$ and $\epsilon_k \sim 0.2$). 
For $L_{rad}=4\times 10^{45}$\ergs\ as measured in Section~\ref{power}, we obtain $3.2\times 10^7$\msun\ for the mass of the black hole powering the FRII galaxy, which for an accretion time of 70 Myr must then have grown from an initial mass $M_0 = 6\times10^6$\msun. Finally, we assumed an ambient gas with uniform 
density of $\rho_{gas}\sim2\times 10^{-28}$ g~cm$^{-3}$ (consistent, e.g., with the average gas density in the $z\sim2$ protocluster studied by \citealt{hennawi15}). By substituting these values in Eqs.15 and 16 of \citet{gilli17}, we find that after 70 Myr, the bubble radius has expanded to 117 kpc, in excellent agreement with the measured size of component A of the diffuse X-ray emission ($\sim 120$ kpc radius). 

We are also now in a position to compare the timescales of the star formation of galaxies $m1-m4$ with the time needed for the shock to cross them. On the one hand, we may expect that star formation is promoted by the compression of the cold gas in the galaxies as the shock approaches them. On the other hand, star formation may be quenched by removal and heating of their ISM (e.g., through ram-pressure stripping and thermal conduction, respectively; \citealt{vollmer01,tonnesen09,vija17}) after the shock has crossed them and they are immersed within the hot ($T>5$ keV) plasma of the bubble. In this scenario the time needed by the shock to cross the galaxies $m1-m4$ should be similar to the age of the burst of star formation producing most of the observed optical and UV emission of $m1-m4$. 
Based on the fit to the UV spectra of galaxies $m1-m4$ with stellar population synthesis models (see Section~\ref{sfragesmasses}), we found
significant episodes of star formation as young as a few megayears in all galaxies. Furthermore, in $m1$ and in $m4$ these recent bursts produce most of the optical and UV light. Using Eq.~18 of \citet{gilli17}, we derived the expansion velocity of the hot gas bubble at t=70 Myr, which is $v=1240$ \kms \ and corresponds to $\sim 1.2$ kpc~Myr$^{-1}$. Galaxies $m1-m4$ appear as compact systems in the HST and LBT/SOUL images. Based on the results obtained with SOUL on $m3$ (and also on the additional candidate radio galaxy with $z_{phot}\approx 1.7$), a typical diameter of 4-5 kpc may be considered for these systems. The time needed by the shock to cross the galaxies is therefore comparable with the age and duration of recent bursts of star formation, which has to be expected if this is truly promoted by the compression of their ISM by the approaching shock front.

To validate this scenario, we verified whether the sSFR, that is, the star formation rate per unit stellar mass, in the galaxies $m1-m4$ is effectively higher than that of galaxies $m5$ and $m6$ and than that of field galaxies of similar mass at the same redshift. In Fig.~\ref{ssfr} ({\it top panel}) we show the SFR versus stellar mass $M_*$ for $m1-m6$ derived in Section~\ref{sfragesmasses} compared with the average relation (the so-called main-sequence) derived by \citet{santini17} for galaxies at $z=1.3-2.0$ in the HST Frontier fields. Among the six MUSE galaxies in the overdensity
$m1-m6$, the four at the bubble boundary, $m1-m4$, have SFRs higher than those expected in main-sequence galaxies, whereas $m5$ and $m6$ are on and below the main sequence, respectively. As shown in the {\it bottom panel} of the same figure, $m1-m4$  feature the highest sSFRs of the six MUSE galaxies at $z=1.7$. 
These rates are a factor of 2-5 higher than the average sSFR of field galaxies of similar mass at z=1.7. Although we cannot make strong statements on the statistical significance of this result, we remark that it is consistent with the expectations of the proposed positive AGN feedback scenario. In the future, observations of UV rest-frame emission lines may be a powerful diagnostics
to confirm the effectiveness of the incoming shock in promoting star formation in these galaxies by compressing and cooling their ISM.

As a final remark, we note that the positive feedback scenario may also hold if the diffuse X-rays seen in component A are produced by IC-CMB rather than by shock-heated gas. 
In this case, a conservative lower limit to the total energy density $U_{tot}$, and hence to the total nonthermal pressure of the lobe $P_{tot}=U_{tot}/3$, can be computed assuming that the
system is in the state of minimum energy density $U_{tot}=(7/3)U_{B}$, where $U_B=B^2/8\pi$ is the magnetic energy density (this state is also often referred to as equipartition because the energies of the relativistic particles and magnetic field are approximately equal). By assuming $B=B_{eq}=5\mu$G (see Sect.~5.3) and the fiducial density and temperatures of the cold ambient medium discussed in this Section, we obtain $P_{tot}/P_{cold}>260$. This suggests again that the lobe is highly overpressurized and hence expands, compressing the ISM of the galaxies at its border and
enhancing their star formation.

\section{Conclusions}

We have reported the discovery of a galaxy overdensity around an FRII radio galaxy at z=1.69 in the multiband survey around the z=6.3 QSO SDSS~J1030+0524. This structure is likely still
far from virialization phase and features significant evidence for positive AGN feedback produced by the radio galaxy. The field features deep multiwavelength coverage from the radio band to the X-rays. We analyzed and discussed observations with VLT/MUSE, LBT/LUCI, LBT/SOUL, VLA, and Chandra. Our main results are summarized below.

$\bullet$ We identified six overdensity members with MUSE and two members with LUCI/LBT. Their spectroscopic redshifts cover the range z=1.687-1.699. 
This structure extends for at least 800 kpc on the plane of the sky (Fig.~\ref{full}) and constitutes a significant (false-detection probability $<3.5\times 10^{-7}$) overdensity of $\delta_g=22$ in redshift space (Fig.~\ref{zdist}). 
Based on the measured overdensity level and structure volume, we derive a lower limit to the total mass of the system of $1-2 \times 10^{13}$\msun.
Most overdensity members are compact and blue galaxies that are forming stars at rates of 8-60 $M_{\odot}$~yr$^{-1}$. We do not observe any strong galaxy 
morphological segregation or concentration around the FRII core. The structure is then populated by gas-rich galaxies and is still far from being virialized. It likely constitutes the progenitor of a local massive galaxy group or cluster caught during its main assembly phase.

$\bullet$ One of the two galaxies studied with LUCI is the host of the FRII, for which a redshift of z=1.699 was measured. The radio structure of the FRII  extends for $\sim$ 600 kpc across the sky. By combining VLA radio data at 1.4GHz with TGSS data at 150 MHz, we derived a kinetic power, velocity, and inclination angle of the FRII jet of $P_{kin}=6.3 \times 10^{45}$ \ergs, $v=0.4-0.5c$ and $\theta=70-80$ deg, respectively.

$\bullet$ Based on a 500ks Chandra ACIS-I observation, we found that the FRII nucleus hosts a luminous QSO ($L_{2-10}=1.3\times10^{44}$ erg~s$^{-1}$, intrinsic and rest-frame) that is obscured by Compton-thick absorption ($N_H=1.5\pm0.6\times 10^{24}$\cm; Fig.~\ref{xid189}). Under standard bolometric corrections the measured disk radiative power ($L_{rad}\sim4 \times 10^{45}$ \ergs) is similar to the jet kinetic power estimated from radio observations,
as is commonly found in powerful jetted AGN.

$\bullet$ We found several regions of diffuse X-ray emission within the structure (see Figs.~\ref{csmooth},~\ref{HR}, and \ref{sepia}). The most significant component (component A) extends for $\sim$240 kpc in diameter around the eastern lobe of the FRII. We investigated the origin of this component and conclude that the most plausible explanation is thermal emission from an expanding bubble of gas that is shock-heated to a temperature of $T\approx 5$ keV by the FRII jet and carries an internal energy of $E_{th}\sim 7\times 10^{60}$ erg. We note that it is difficult to accurately measure the gas energy and temperature because of some possible IC-CMB contribution to the X-ray emission. However,  the estimated values can be easily explained by the measured jet power and velocity if the system lifetime is  $\sim$70 Myr, which is typical of FRIIs. By means of a simple model for the expansion of hot gas bubbles inflated by exponentially growing supermassive black holes, we verified that the measured radius of $\sim 120$ kpc of the diffuse X-ray component A is in excellent agreement with the expected radius given the estimated jet power and lifetime.

$\bullet$ Four ($m1-m4$) out of the six MUSE star-forming galaxies in the overdensity are distributed in an arc-like shape at the edge of the main component
(component A) of the diffuse X-ray emission. These four objects are concentrated within 200 kpc in the plane of the sky and within 450 kpc in radial separation. Three of them are even more concentrated, falling with 60 kpc in both transverse and radial distance. The probability of observing four out of these six actively star-forming galaxies at the edge of the diffuse emission by chance is negligible. We then propose that star formation in these galaxies is promoted by the compression of their ISM by the approaching shock front, which would be remarkable evidence of positive AGN feedback on cosmological scales. The sSFRs of $m1-m4$ are the highest of the six MUSE overdensity members (see Fig.~13) and lie above the average sSFR of field galaxies with similar mass at z=1.7, in agreement with the proposed positive AGN feedback scenario. We emphasize that our conclusions about the feedback are robust because even when we assume that all of the diffuse X-ray emission is due to IC-CMB, star formation can still be promoted by the nonthermal pressure of the expanding lobe.\\

In the near future, several multiwavelength observational campaigns will bring further insight into the physics and structure of the overdensity.
As an example, we recently observed the whole region around the FRII radio galaxy with a small ALMA mosaic to reveal a possible molecular 
gas reservoir around the FRII host and discover additional cluster galaxy members through the detection of the CO(2-1) emission line. 
In addition, we observed the field at 1.4GHz with 
the Jansky Very Large Array down to $\sim 1.8\mu$Jy $rms$, that is, one dex deeper than what is currently available, and the newly allocated LOFAR  observations will improve the angular resolution at 150MHz by a factor of three with respect to the existing TGSS data. The new radio datasets are expected to probe the morphology
of the radio structure and reveal its interplay with the observed diffuse X-ray emission with an unprecedented level of detail for a distant large-scale structure.

\begin{acknowledgements}

We wish to thank the referee for a careful and critical review of the paper that helped us to present our results in a more balanced way.
We acknowledge the support from the LBT-Italian Coordination Facility for the execution of observations, data distribution and reduction. The LBT is an international collaboration among institutions in the United States, Italy and Germany. LBT Corporation partners are the University of Arizona on behalf of the Arizona university system; Istituto Nazionale di Astrofisica, Italy; LBT Beteiligungsgesellschaft, Germany, representing the Max-Planck Society, the Astrophysical Institute Potsdam, and Heidelberg University; The Ohio State University, and The Research Corporation, on behalf of The University of Notre Dame, University of Minnesota and University of Virginia. The scientific results reported in this article are partly based on observations made by the Chandra X-ray Observatory. This work made use of
data taken under the ESO program ID 095.A-0714(A).
We acknowledge useful discussions with L. Pozzetti and S. Ettori. We thank P. Santini for providing the literature data points of Fig.~13.
We acknowledge financial contribution from the agreement ASI-INAF n. 2017-14-H.O. IP and QD acknowledge support from INAF under the PRIN SKA project FORECaST (1.05.01.88.03).
\end{acknowledgements}

\bibliographystyle{aa} 

\bibliography{/Users/gilli/protocluster/biblio}

\end{document}